\documentclass[showpacs,prb,twocolumn,floats,superscriptaddress]{revtex4}

\usepackage{graphicx}
\usepackage{amsmath}
\usepackage{amssymb}
\usepackage[colorlinks=true,citecolor=blue,linkcolor=blue]{hyperref}
\usepackage{amsfonts}
\usepackage[latin9]{inputenc}
\usepackage{color}
\usepackage{textcomp}
\usepackage{multirow}
\usepackage{makecell}

\usepackage{placeins}
\renewcommand{\vec}[1]{\ensuremath{\boldsymbol{#1}}}

\begin{document}

\title{Confined states in graphene quantum blisters}
\date{\today }
\author{H. M. Abdullah}
\email{alshehab211@gmail.com}
\affiliation{Department of Physics, King Fahd University of Petroleum and Minerals, 31261 Dhahran, Saudi Arabia}
\affiliation{Saudi Center for Theoretical Physics, P.O. Box 32741, Jeddah 21438, Saudi Arabia}
\affiliation{Department of Physics, University of Antwerp, Groenenborgerlaan 171, B-2020 Antwerp, Belgium}
\author{H. Bahlouli}
%\email{Bahlouli@kfupm.edu.sa}
\affiliation{Department of Physics, King Fahd University of Petroleum and Minerals, 31261 Dhahran, Saudi Arabia}
\affiliation{Saudi Center for Theoretical Physics, P.O. Box 32741, Jeddah 21438, Saudi Arabia}
\author{F. M. Peeters}
%\email{francois.peeters@uantwerpen.be}
\affiliation{Department of Physics, University of Antwerp, Groenenborgerlaan 171, B-2020 Antwerp, Belgium}

\author{B. Van Duppen}
\email{ben.vanduppen@uantwerpen.be}
\affiliation{Department of Physics, University of Antwerp, Groenenborgerlaan 171, B-2020 Antwerp, Belgium}

\begin{abstract}
Bilayer graphene samples may exhibit regions where the two layers are locally delaminated  forming a so-called quantum blister in the graphene sheet. Electron and hole states can be confined  in this graphene quantum blisters (GQB) by applying  a global electrostatic bias.  We scrutinize the electronic properties of  these confined states  under the variation of interlayer bias, coupling, and blister's size.   The spectra display strong anti-crossings due to the coupling of the confined states on upper and lower layers inside the blister. These spectra   are layer localized where the respective confined states reside on either  layer or equally distributed. For finite angular momentum, this layer  localization can be at the edge of the blister and corresponds to degenerate modes of opposite momenta.  Furthermore, the energy levels in  GQB    exhibit electron-hole symmetry that is sensitive to the electrostatic bias.    Finally, we demonstrate that confinement in GQB persists even in the presence  of a variation in the inter-layer coupling.

\end{abstract}

\maketitle

%%%%%%%%%%%%%%%%%%%%%%%%%%%%%%%%%%%%%%%%%%%%
\section{Introduction}
%%%%%%%%%%%%%%%%%%%%%%%%%%%%%%%%%%%%%%%%%%%%

In its natural form, graphite consists of many layers of carbon atoms stacked in a Bernal configuration\cite{Bernal_1924}. When scaling down such graphitic systems to the few-layers level, it is possible that perfect Bernal stacking is broken. For example two neighbouring layers can be shifted\cite{Daboussi2014,Gradinar2012} or rotated\cite{Gail2011,Lopes_dos_Santos_2012,Lopes_dos_Santos_2007,Mele2012,Rong1993} as a whole with respect to the Bernal configuration. Other systems can exhibit local transitions from an AB-stacking to BA-stacking of two layers resulting in stacking domain walls\cite{Yin2016}. Each of these structural deformations result in distinct changes to the electronic properties of few-layered systems.

Recently, another type of deformation was observed in bilayer graphene. In this case, the two layers are connected in an usual Bernal configuration, but locally depart from each other\cite{Hao2016,Wang2017}. It was noted previously that the formation of these kinds of structures has a strong impact on the transport properties of charge carriers in bilayer graphene systems\cite{Abdullah2017}. As a consequence, it was proposed to use these kinds of deformations to make devices that are layer-selective. \cite{Abdullah_2016,Lima2016,Brandimarte2017}. 

\begin{figure}[t!]
\vspace{0.cm}
\centering\graphicspath{{./Figures/}}
\includegraphics[width=1.96  in]{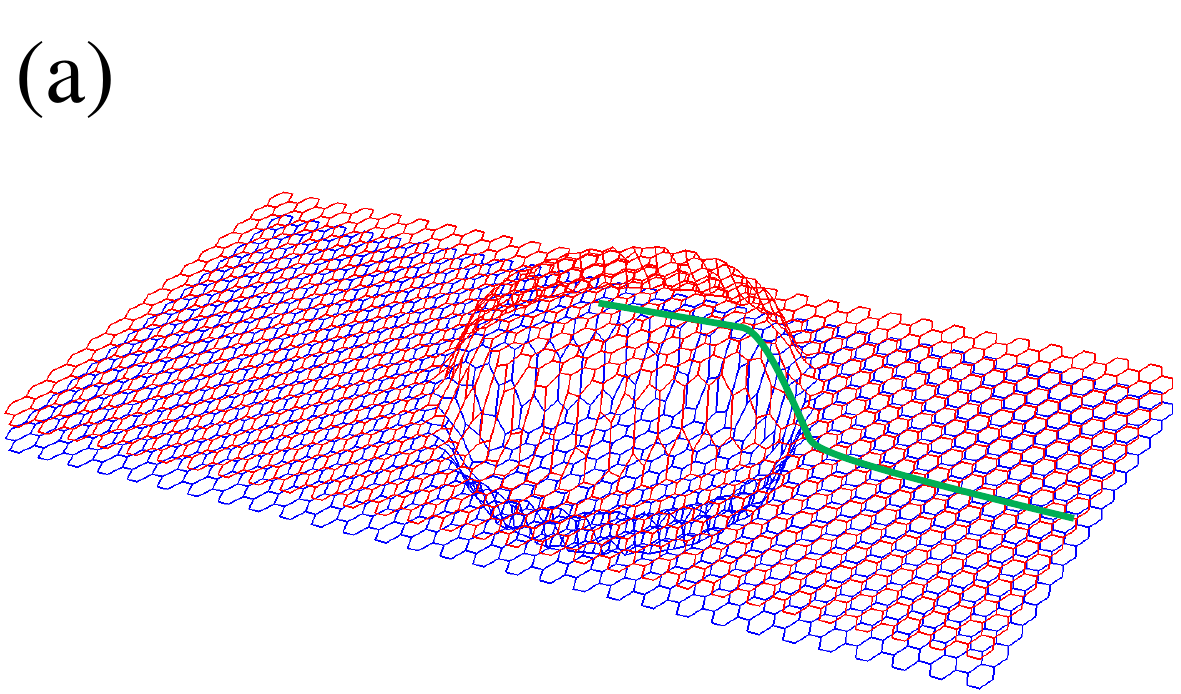}\
\includegraphics[width=1.3  in]{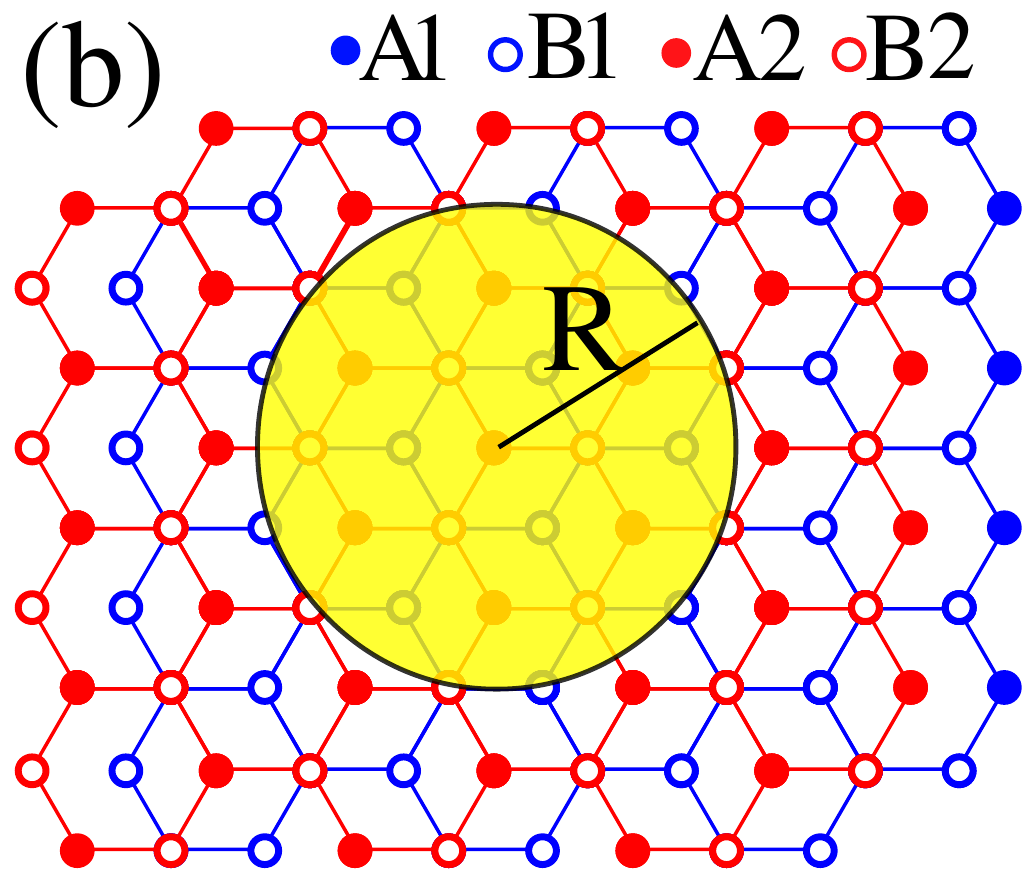}\\ \ \ \ \ \ \ \ \ \
\includegraphics[width=2.7  in]{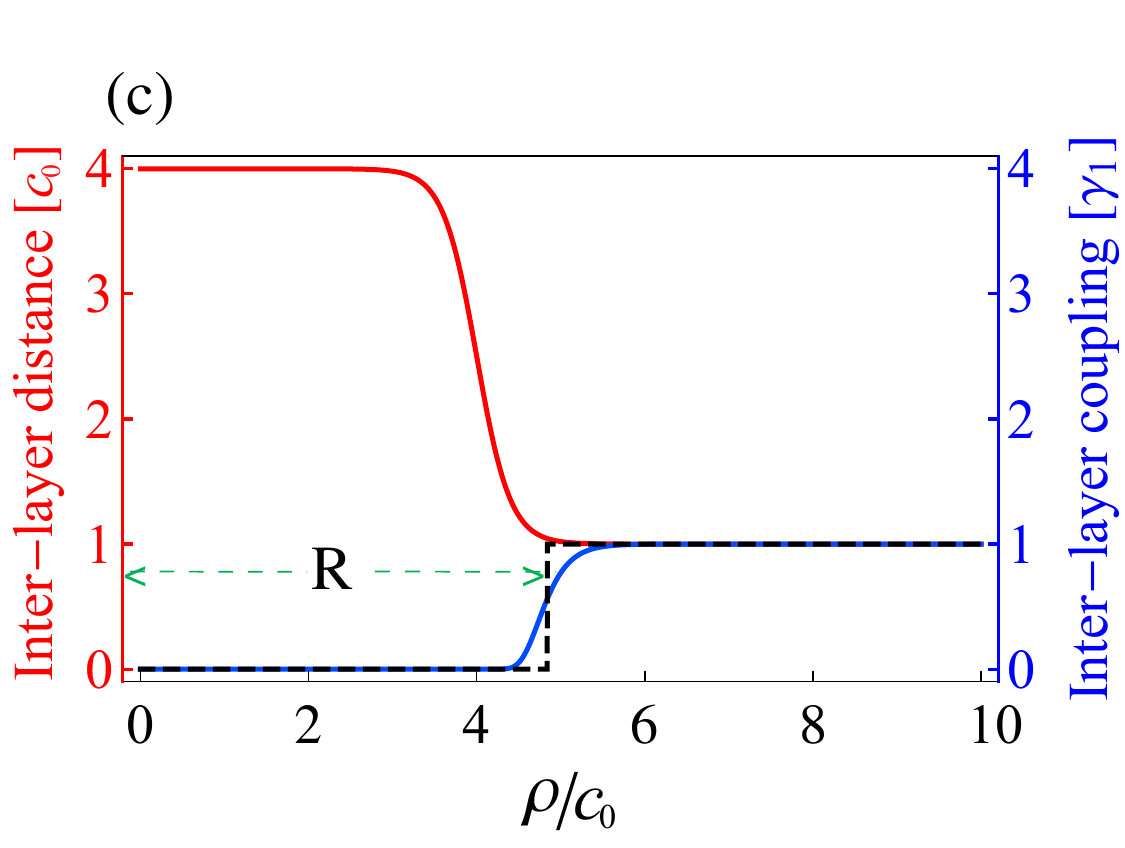}\\
\includegraphics[width=2.7  in]{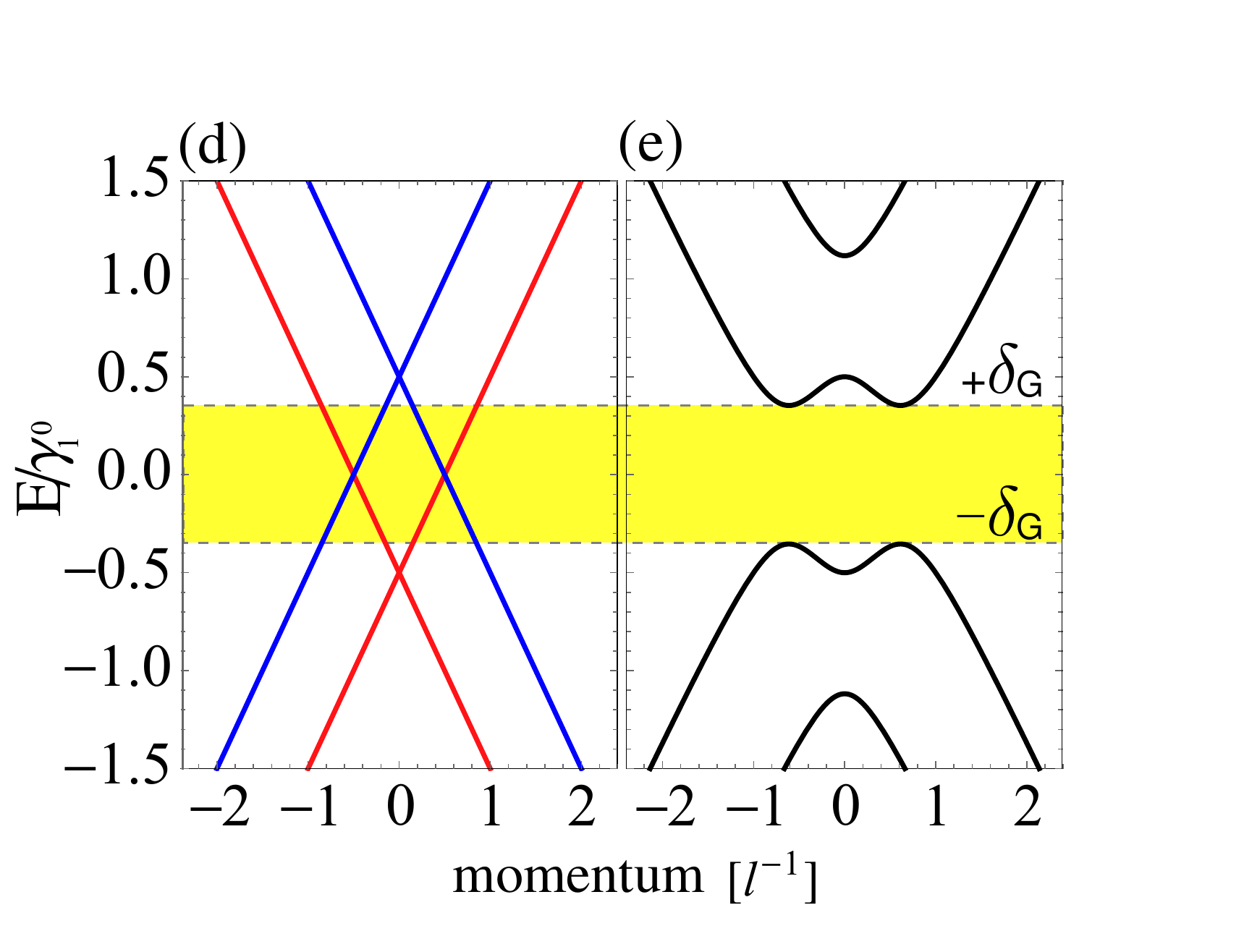}\\
\caption{(Color online)  (a) Schematic pictures of the proposed circular GQB with radius R with Bernal stacking outside the blister as shown in panel (b). (c)  Inter-layer distance and coupling as a function of $\rho$ along  the green curve in (a), the dashed black curve represents the abrupt change in the inter-layer coupling at the interface R of the GQB. (d, e) energy spectrum inside and outside the GQB in  the presence of a global interlayer bias $\delta$, respectively.  Red and blue bands in (d) correspond to upper and lower layers inside the blister and $\gamma_{1}^{0}=0.38$ eV is the standard inter-layer coupling in AB-stacked bilayer graphene.  The confinement is possible in the energy range $E<\left\vert \delta_{G} \right \vert$   delimited by the yellow region,  where $\delta_{G}$ represents the direct energy gap  defined in  Eq.\eqref{eq_energy_range}.} \label{fig-GQB}
\end{figure}

The work on electronic transport mentioned above considered a deformation of infinite length across which current flow was studied. However, it is also possible that the region where the two graphene layers are departing from each other is finite in size. These defects occur naturally\cite{Yan2016,Schmitz2017a,Clark_2014}, or one can imagine that such structure can be created by intercalating nano-clusters of atoms in-between the neighbouring sheets\cite{Kim2011a} or even deliberately grow such structures on graphene sheets decorated with nanostructures\cite{Scheerder2017}. The basic structure of a blisters is depicted in Fig. \ref{fig-GQB}(a) and is called a graphene quantum blister (GQB)\cite{Abdullah2018}. 

 A rigorous control of the charge carriers is indispensable in electronic devices fabrication. Essentially, this requires    a perfect confinement  of the charge carriers as well as external tunability. Unfortunately, perfect electrostatic confinement in graphene is precluded    by Klein tunneling and only quasi-confined states  with a finite trapping lifetime are allowed\cite{Matulis2008} or by applying a uniform magnetic filed\cite{Giavaras2012}.  Apart from the electrostatic confinement, different routs have been proposed to trap electrons in single layer graphene. For example, introducing a gap induced by the substrate\cite{Giavaras2011,Giavaras2010,Recher2009} or cutting a graphene flake into small areas\cite{Hewageegana2008,Costa2014,Mirzakhani2016,Zarenia2010,Zebrowski2013}. However, these proposal associate with some experimental dificulties such as the high sensitivity of QDs to
their precise terminations and the control of the induced gap by the dielectric substrates.   This hampered deployment of graphene in electronic devices.   Recently, many experiments\cite{Zhao2015,Ghahari2017,Gutierrez2016,Lee2016,Freitag2016} have  realized such quasi-confined states in quantum dots (QDs) through different approaches. For example, by  employing the   electrostatic potential induced by  the tip of the scanning tunneling microscope (STM) \cite{Zhao2015} or in the presence of a strong magnetic field\cite{Freitag2016}. A GQB not only supports electrostatic confined states with infinite  trapping lifetime but also allows external electrostatic tunability. Moreover, its electronic properties are  amenable to direct STM measurements\cite{Morgenstern2017}. 

An alternative way followed in the quest for electronic confinement in graphene uses bilayer graphene as a basis \cite{Mueller2014,J.MiltonPereira2007,Zarenia2009}. For these systems the electronic energy  spectrum is parabolic instead of linear as is the case  in single layer graphene, and can be gapped by applying a potential bias to the two layers \cite{McCann2006}. By nano-engineering electrostatic gates such that the bilayer graphene spectrum is gapped everywhere except in a locally defined region, charge carriers can indeed be confined \cite{J.MiltonPereira2007}. However, in practice it is challenging to engineer the gates such that the bias profile has the desired shape and the resulting confined electron states loose their interesting ultra-relativistic character. 

 A GQB is a peculiar system, especially in the presence of an external electric field. Indeed, the two graphene layers are nearly delaminated in the center of the blister, while they are composed into Bernal stacking outside of the GQB. Both a delaminated graphene layers     and a Bernal bilayer have gapless energy spectra with  massless Dirac Fermions for the former and massive Fermions for the latter\cite{Goerbig_2011,Rozhkov_2016,Castro_Neto_2009}. However, the response to an external electric field that results from the application of a potential difference between the two layers is fundamentally different for both systems. While for the delaminated layers the Dirac cones of each layer shift in energy, see Fig. \ref{fig-GQB}(d), a gap is opened in the Bernal bilayer spectrum \cite{Zhang_2009,Ohta_2006} as shown in Fig. \ref{fig-GQB}(e). As a consequence, electronic states with energy in the gap have to be confined in the vicinity of the blister. As such, the GQB becomes a quantum dot with a discrete energy spectrum.

Recently, this GQB system\cite{Abdullah2018} has been introduced  to achieve  an ideal electrostatic confinement  in a delaminated bilayer graphene of a Gaussian-dome shape. It also showed the ability of  controlling the layer localization by an electrostatic gate for zero angular momentum.   Here we consider the blister as  delaminated bilayer graphene, which can be considered as monolayer-like system, that connected to AB-stacked bilayer graphene through an  abrupt interface, see Fig. \ref{fig-GQB}(a). We systematically scrutinize  the electronic properties of the confined states by inspecting the effect of different parameters such as inter-layer bias and coupling as well as size of the blister. Furthermore, within the four band Hamiltonian we present an analytical model to calculate the wave functions in GQB and the respective confined states as well as the local density of state.   

The characteristics of
confined modes in GQBs are mainly sensitive to the global bias. Of particular importance, in the case of  homogeneous bias, the confined modes are layer dependent and inherit the symmetry $E_{m,n}=E_{-m,n}$ where $m$ and $n$ are  the angular and radial quantum numbers, respectively. On the other hand, considering pristine blisters or with non-homogenous  bias leads to a different symmetry, namely,   $E_{m,n}=-E_{-m,n}$ and the latter introduces degenerate modes  at zero energy for non-zero angular momenta.  Such modes live at the interface of the blister and are localised on different layers.   Finally,  the obtained   energy spectrum is found to be  robust with respect to changes in the inter-layer coupling inside the GQB and therefore the results obtained in this paper are expected to be widely visible in experiments.

The paper is structured as follows. In Sec. \ref{Sec:Model}, we discuss how confinement can be realized and   present the  electronic model to calculate the bound states in GQB. In Sec. \ref{Results}, we investigate the effect of homogeneous and non-homogeneous inter-layer bias and discuss the character of the confined states. Finally,  in Sec. \ref{Conclusions}, we  draw our conclusions and highlight the main findings.

%%%%%%%%%%%%%%%%%%%%%%%%%%%%%%%%%%%%%%%%%%%%
\section{Electronic model}\label{Sec:Model}
%%%%%%%%%%%%%%%%%%%%%%%%%%%%%%%%%%%%%%%%%%%%
\subsection{Electrons in bilayer graphene}
In Figs. \ref{fig-GQB}(a,b), we show schematically the atomic structure of a circular GQB with radius $R$. In the continuum limit, \textcolor{red}{} if the size of the GQB is much larger than the inter-atomic distance, one can describe charge carriers by a $4\times4$ tight-binding Hamiltonian written in the basis of orbital eigenfunctions of the four atoms making up the crystallographic unit cell of bilayer graphene\cite{McCann2006}. The labelling of the different atoms is shown in Fig. \ref{fig-GQB}(b). In the basis $\mathbf{\Psi}=(\Psi_{A 1},\Psi_{B 1},\Psi_{B 2},\Psi_{A 2})^{T}, $  the Hamiltonian in position representation in valley $K^{\tau}$ reads
\begin{equation}\label{starting_Hamiltonian}
\hat{H}^\tau(\vec{r})=\left(
\begin{array}{cccc}
   \tau \delta  & v_{\rm F}\hat{\pi}_{+}^{\tau} &  \gamma_{1}(\vec{r}) & 0 \\
  v_{\rm F}\hat{\pi}^{\tau}_{-} & \tau \delta &  0 & 0\\
  \gamma_{1}(\vec{r}) &   0& - \tau \delta & v_{\rm F}\hat{\pi}^{\tau}_{-} \\
  0 & 0& v_{\rm F}\hat{\pi}^{\tau}_{+} & - \tau \delta  \\
\end{array}%
\right)~.
\end{equation}
In Eq. (\ref{starting_Hamiltonian}), $v_{\rm F}\sim 10^6~{\rm m/s}$ is the graphene Fermi velocity\cite{Castro_Neto_2009}  and $\tau=(+1,-1)$ for the $K-$ and $K'-$valley, respectively. Furthermore, we have neglected skew hopping parameters that only affect the energy spectrum for very small energies\cite{Van_Duppen01_2013}  and only retain $\gamma_{1}(\vec{r})$, the inter-layer coupling through van der Waals forces\cite{Li2009}. The quantity $\delta$ denotes the potential bias between the two layers and   $\hat{\pi}_{\pm}^{\tau}$ are the momentum operators, which in polar coordinates become \cite{Mirzakhani2016} 
\begin{equation}\label{momentum_operator}
\hat{\pi}_{\pm}^{\tau}=\frac{\hbar}{i}
e^{\pm i\tau \phi}\left[ \frac{\partial}{\partial \rho} \pm \frac{i \tau}{\rho}\frac{\partial}{\partial \phi} \right]~.
\end{equation}
In Eq. \eqref{momentum_operator}, $\rho$ is the radial distance to the center of the blister and $\phi$ is the azimuthal angle. Notice that the momentum operator depends on the valley in which the charge carriers reside\cite{da_Costa_2015,Rycerz2007}. 

\subsection{Inter-layer coupling in a GQB}

In Eq. \eqref{starting_Hamiltonian} the function $\gamma_{1}(\vec{r})$ describes the coupling between the two graphene layers. If we consider the GQB as in Fig. \ref{fig-GQB}(a) where the layers depart from each other  in the form of a  kink, one can describe the inter-layer distance $c(\vec{r})$ as
\begin{equation}
c(\vec{r}) = \frac{(c_{\rm M}-c_0)}{2} \left[ \tanh\left(\frac{-\rho+R_{\rm QB}}{\xi}\right)+3 \right]~,
\label{inter_layer_distance}
\end{equation}
where $c_{0} \sim 0.33~{\rm nm}$  is the equilibrium inter-layer distance, $c_{\rm M}$ is the inter-layer distance at the center of the blister, $R_{\rm QB}$ is the radius of the blister, $\xi$ is the width of the  interface  between  delaminated and AB-stacked bilayer graphene,  and $\rho$ is the radial component. Because the inter-layer coupling strength $\gamma_{1}(\vec{r})$ arises from the overlap of two orbital eigenfunctions in the tight-binding formalism, its value decreases exponentially with increasing inter-layer distance. Following standard practice\cite{Lopes_dos_Santos_2007,Lopes_dos_Santos_2012,Donck2016}, we can write the inter-layer coupling function as
\begin{equation}
\gamma_{1}(\vec{r}) = \gamma^{0}_{1} \exp\left(-\beta \frac{c(\vec{r})-c_{0}}{c_{0}}\right)~.
\label{gamma1_coupling}
\end{equation}
In Eq. \eqref{gamma1_coupling} we have introduced    $\gamma_{1}^{0} = 0.38~{\rm eV}$ \cite{Xu_2010,Lobato_2011,Li2009} the equilibrium inter-layer coupling. The quantity $\beta/c_{0}$ is the inverse inter-layer coupling decay length. For the calculations in this paper we choose $\beta \sim 13.3$, as was used before to match with the values for the skew hopping parameters in twisted bilayer graphene\cite{Donck2016}. However, this value can be even larger when the blister is formed by insertion of nanoclusters in-between two graphene layers as these clusters screen the van der Waals interaction between the layers. 

By plugging Eq. \eqref{inter_layer_distance} into Eq. \eqref{gamma1_coupling}, one can calculate the radial dependence of the inter-layer coupling for a GQB. In Fig. \ref{fig-GQB}(c) we compare the lateral profile of the GQB to the strength of the inter-layer coupling at the same positions. Because the magnitude of the inter-layer coupling decreases exponentially with inter-layer distance, already for a very small change in inter-layer distance, the coupling is diminished. Outside the GQB, however, the coupling still attains the value $\gamma_{1}^{0}$. As a result, to a very good approximation it is safe to assume that the two graphene layers are decoupled for the entire size of GQB, while just outside the blister the layers are coupled. Hence,  we assume in this work that   the inter-layer coupling profile has an abrupt transition at position $R$ as shown in Fig. \ref{fig-GQB}(c) by the black-dashed line. However, the results were shown to be robust against  a smoothed blister\cite{Abdullah2018}.
\subsection{Electronic confinement}

In Figs. \ref{fig-GQB}(d,e) we show the energy spectra, respectively, inside and outside the GQB under the application of a finite inter-layer bias $\delta$. Because of the lack of inter-layer coupling the energy spectrum is linear and gapless inside the GQB. The application of a different potential to both layers, therefore, shifts the Dirac point in energy. As a result, for every energy there are electron or hole states available. Outside the GQB, Fig. \ref{fig-GQB}(e) shows that the situation is substantially different. Because here the inter-layer coupling is strong, the inter-layer bias $\delta$ opens up a gap in the energy spectrum. In this region, only evanescent states are allowed and, therefore, the energy spectrum inside the GQB will be discrete and the corresponding modes are confined. The energy range where confinement appears is given by the range $\left[-\delta_{\rm G},\delta_{\rm G}\right]$, where $\delta_{\rm G}$ is related to the inter-layer bias as\cite{Abdullah_2017}
\begin{equation}
\delta_{\rm G} = \delta \left(1+4\frac{\delta^{2}}{\gamma_{1}^{2}}\right)^{-1/2}~.
\label{eq_energy_range}
\end{equation}

In the following section we calculate the wavefunctions and the respective  energy spectrum  of the confined states in a GQB with an abrupt interface. For this, we first need to obtain the wavefunctions inside and outside the GQB and require continuity for each of the spinor components at the interface  $\rho = R$ to find the energy eigenstates of the GQB in the presence of an inter-layer bias. In all calculations and results, the energy is scaled with the equilibrium inter-layer hopping parameter, $\gamma_1^{0}$, while $l = \hbar v_{\rm F} / \gamma_{1}^{0} \sim 1.65~{\rm nm}$ is the measure for the length scales. 

\subsection{Wavefunctions outside the GQB} 
In order to obtain the wavefunction outside the GQB, we solve the Schr\"{o}dinger equation $\hat{H}^{\tau}(\vec{r}) \Phi^{\tau}(\vec{r}) = E \Phi^{\tau}(\vec{r})$ for the Hamiltonian given in Eq. \eqref{starting_Hamiltonian} with $\gamma_{1}(\vec{r}) = \gamma_{1}^{0}$. The equation for the angle $\phi$ directly yields a relation between the phases of each spinor component. This means that the four-component wave function $\Phi^{\tau}(\vec{r})$ in the $\tau$ valley can be written as\cite{Mirzakhani2016}
\begin{equation}\label{wave_spinor_angular}
 \mathbf{\Phi^{\tau}}(\vec{r})=\left(
\begin{array}{cccc}
  \phi_{A1}^{\tau}(\rho)e^{im\phi}\\
   i\phi_{B1}^{\tau}(\rho)e^{i(m-\tau)\phi}\\
    \phi_{B2}^{\tau}(\rho)e^{im\phi}\\
   i\phi_{A2}^{\tau}(\rho)e^{i(m+\tau)\phi}
\end{array}
\right)~.
\end{equation}
Solving the Schr\"odinger equation for the radial functions $\phi^{\tau}_{i}(\rho)$, we obtain the following set of coupled equations:
\begin{subequations}
\begin{eqnarray}
&&\left[\frac{d}{d\rho}-\frac{(\tau m-1)}{\rho}\right]\phi_{B1}^{\tau}=(E-\tau\delta)\phi_{A1}^{\tau}-\phi_{B2}^{\tau}~,\label{eq-06}\\
&&
\left[\frac{d}{d\rho}+\frac{\tau m}{\rho}\right]\phi_{A1}^{\tau}=-(E-\tau\delta)\phi_{B1}^{\tau}~,\label{eq-07}\\
 &&\left[\frac{d}{d\rho}+\frac{(\tau m+1)}{\rho}\right]\phi_{A2}^{\tau}=(E+\tau\delta)\phi_{B2}^{\tau}-\phi_{A1}^{\tau}~,\label{eq-08}\\
&&\left[\frac{d}{d\rho}-\frac{\tau m}{\rho}\right]\phi_{B2}^{\tau}=-(E+\tau\delta)\phi_{A2}^{\tau}~.\label{eq-09}
\end{eqnarray}
\end{subequations}

We remind the reader that in this set of equations, the energetic quantities are scaled with $\gamma_{1}^{0}$ and the radial component $\rho$ by $l$, yielding dimensionless equations. The set of first-order differential equations can be written as a single fourth-order differential equation. As explained  previously\cite{Xavier2010}, this fourth-order differential equation has two sets of orthogonal solutions given by the solutions of the following second-order differential equations: 
\begin{equation}\label{eq-10}
\left[\frac{d^{2}}{d\rho^{2}}+\frac{1}{\rho}\frac{d}{d\rho}-\left( \frac{m^{2}}{\rho^{2}}+\alpha_{\pm}^{2} \right)\right]\phi_{A1}^{\tau}(\rho)=0~.
\end{equation}
The two equations only differ by the value of
\begin{equation}
\alpha_{\pm}^{2}=-(E^{2}+\delta^{2})\pm\sqrt{(E^{2}-\delta^{2})+4E^{2}\delta^{2}}~.
\label{alpha_equation}
\end{equation}
In the energy range where confinement is expected, the square root of Eq. \eqref{alpha_equation} is imaginary. As a consequence, the solutions to Eq. \eqref{eq-10} are Bessel functions with a complex argument\cite{abramowitz1964handbook}. Because we are outside of the GQB, the spinor components need to be finite in the limit $\rho\rightarrow \infty$, so we choose the modified Bessel function of the second kind $K_{m}(\alpha_{\pm} \rho)$ as solutions. Finally, notice that $\alpha_{+} = \alpha_{-}^{*}$, such that the two independent solutions of Eq. \eqref{eq-10} can be written as a superposition of the real and imaginary part of $K_{m}(\alpha_{\pm} \rho)$, and we have
\begin{subequations}
\begin{equation}\label{eq-11}
\phi_{A1}^{\tau}(\rho)=C_{1}^{\tau}\Re \left[ K_{m}(\alpha_{+} \rho) \right]+C_{2}^{\tau}\Im \left[ K_{m}(\alpha_{-} \rho) \right],
\end{equation}
Using Eqs.(\ref{eq-06}-\ref{eq-09}) we can obtain the other components explicitly as
\begin{widetext}
\begin{equation}\label{eq-12}
\phi_{B1}^{\tau}(\rho)=\frac{1}{(E-\tau \delta)}\left( C_{1}^{\tau}\Re \left[\alpha_{+} K_{m-\tau}(\alpha_{+} \rho) \right]+C_{2}^{\tau}\Im \left[\alpha_{-} K_{m-\tau}(\alpha_{-} \rho) \right] \right)~,
\end{equation}
\begin{equation}\label{eq-13}
\phi_{B2}^{\tau}(\rho)=\frac{1}{(E-\tau \delta)}\left( C_{1}^{\tau}\Re \left[\eta^{+} K_{m}(\alpha_{+} \rho) \right]+C_{2}^{\tau}\Im \left[\eta^{-} K_{m}(\alpha_{-} \rho) \right] \right)~,
\end{equation}
\begin{equation}\label{eq-14}
\phi_{A2}^{\tau}(\rho)=\frac{1}{(E^{2}- \delta^{2})}\left( C_{1}^{\tau}\Re \left[\eta^{+}\alpha_{+} K_{m+\tau}(\alpha_{+} \rho) \right]+ C_{2}^{\tau}\Im \left[\eta^{-}\alpha_{-} K_{m+\tau}(\alpha_{-} \rho) \right] \right)~.
\end{equation}
\end{widetext}
\end{subequations}

In these equations, we have introduced the compact notation $\eta^{\pm}=\alpha_{\pm}^{2}+(E-\tau \delta)^{2}$.  

\begin{figure}[t!]
\vspace{0.cm}
\centering\graphicspath{{./Figures/}}
\includegraphics[width=3.5 in]{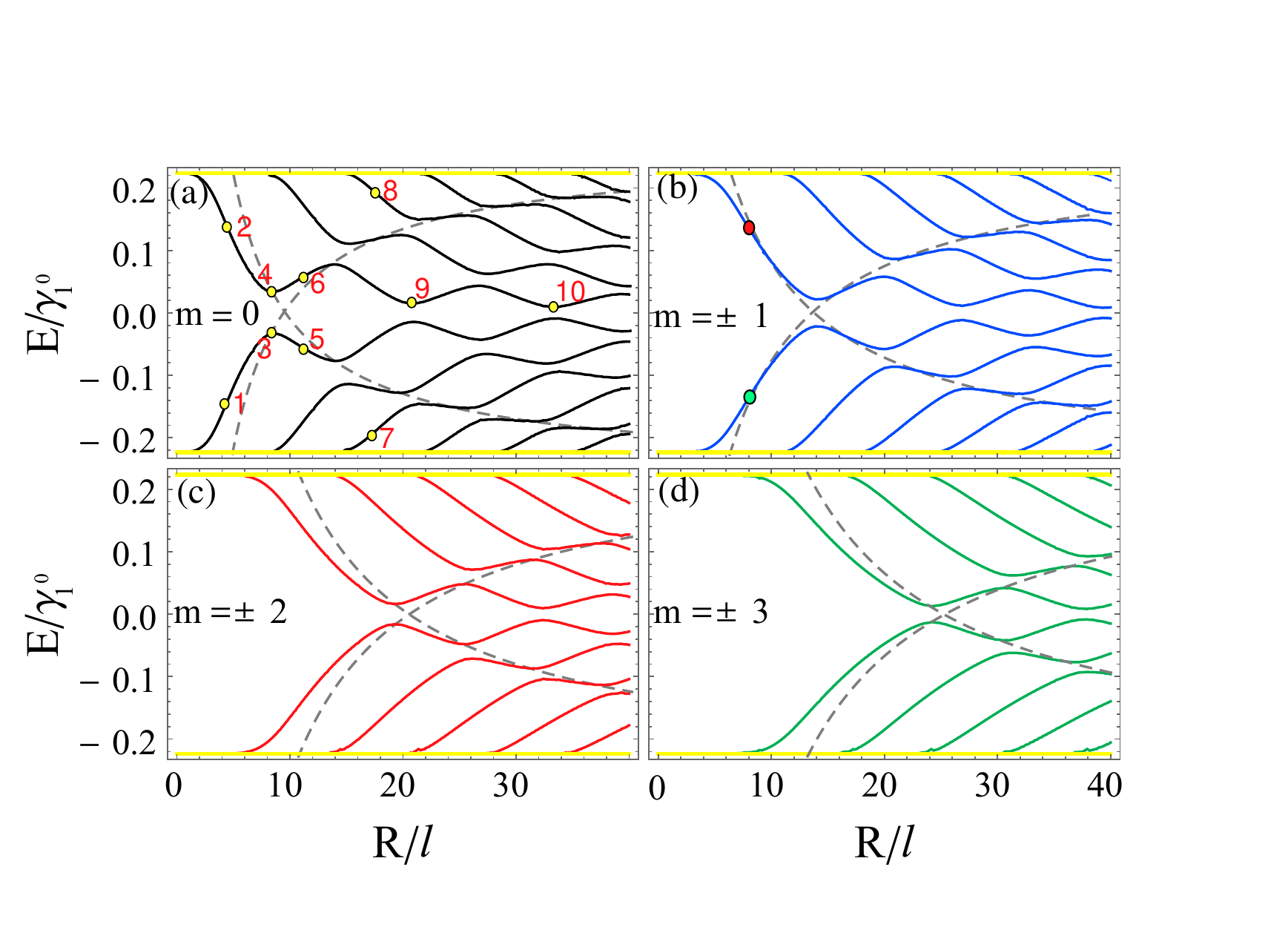}\
\vspace{0.cm}
\caption{Energy levels of the GQB as a function of its radius with the same bias inside and outside the GQB  $\delta_{<}=\delta_{>}=0.25 \gamma_{1}$. Gray dashed curves  correspond to the first energy levels of  a biased bilayer nano-disk.}\label{GQB_energy_levels_same_bias}
\end{figure}
\subsection{Wavefunctions inside the GQB}
Inside the GQB the inter-layer coupling vanishes and, therefore, in Eq. \eqref{starting_Hamiltonian} we have to put $\gamma_{1}(\vec{r})=0$. Although the angular solution of the Schr\"{o}dinger equation stays the same as in Eq. \eqref{wave_spinor_angular}, the set of radial equations changes to
\begin{subequations} 
\begin{eqnarray}
&&\left[\frac{d}{d\rho}-\frac{(\tau m-1)}{\rho}\right]\phi_{B1}^{\tau}=(E-\tau\delta)\phi_{A1}^{\tau}\label{eq-16}~,\\
&&
\left[\frac{d}{d\rho}+\frac{\tau m}{\rho}\right]\phi_{A1}^{\tau}=-(E-\tau\delta)\phi_{B1}^{\tau}\label{eq-17}~,\\
 &&\left[\frac{d}{d\rho}+\frac{(\tau m+1)}{\rho}\right]\phi_{A2}^{\tau}=(E+\tau\delta)\phi_{B2}^{\tau}\label{eq-18}~,\\
&&\left[\frac{d}{d\rho}-\frac{\tau m}{\rho}\right]\phi_{B2}^{\tau}=-(E+\tau\delta)\phi_{A2}^{\tau}\label{eq-19}~.
\end{eqnarray}
\end{subequations} 
In this case, the set of equations is already decoupled for each layer. This allows to find a second-order differential equation for each layer as
\begin{equation}\label{eq-20}
\left[\frac{d^{2}}{d\rho^{2}}+\frac{1}{\rho}\frac{d}{d\rho}-\left( \frac{m^{2}}{\rho^{2}}-\mu_{\pm}^{2} \right)\right]\phi_{B2/A1}^{\tau}(\rho)=0.
\end{equation}
In Eq. \eqref{eq-20}, $\mu_{\pm}=E\pm \tau\delta$ and the subscript of the function $\phi_{i}^{\tau}(\rho)$ refers to $B1$ for $\mu_{+}$ and to $A1$ for $\mu_{-}$. The solutions of Eq. \eqref{eq-20} are Bessel functions. Dropping the ones singular at the origin, we find
\begin{subequations}
\begin{equation}\label{eq-21}
\phi_{A1}^{\tau}(\rho)=D_{1}^{\tau}  J_{m}(\mu_{-} \rho)~,
\end{equation} 
and 
\begin{equation}\label{eq-23}
\phi_{B2}^{\tau}(\rho)=D_{2}^{\tau}  J_{m}(\mu_{+} \rho)~.
\end{equation}
\end{subequations}
The other two components can then be found from Eqs. \eqref{eq-17} and \eqref{eq-19} and yield
\begin{subequations} 
\begin{equation}\label{eq-22}
\phi_{B1}^{\tau}(\rho)=-\tau D_{1}^{\tau}  J_{m-\tau}(\mu_{-} \rho)~,
\end{equation}
and
\begin{equation}\label{eq-24}
\phi_{A2}^{\tau}(\rho)=\tau D_{2}^{\tau}  J_{m+\tau}(\mu_{+} \rho)~.
\end{equation}
\end{subequations}\\
We are now in a position to find the eigenstates and energylevels of a GQB. For this, we need to equate the spinor components inside the GQB with those outside at position $\rho = R$. Doing so, one obtains a set of four equations for four unknowns that can be written in a matrix formalism as
\begin{widetext}
\begin{equation}\label{eq-05}
\mathbf{M^{\tau}}\left(
\begin{array}{cccc}
  D_{1}^{\tau}\\
   D_{2}^{\tau}\\
    C_{1}^{\tau}\\
  C_{2}^{\tau}
\end{array}
\right)=\left(
\begin{array}{cccc}
  -J_{m}(R\mu_{-}) & 0 & \Re[K_{m}(R\alpha_{+})] & \Im[K_{m}(R\alpha_{-})] \\
  \tau J_{m-\tau}(R\mu_{-}) & 0 &  \Re[b_+K_{m-\tau}(R\alpha_{+})] & \Im[b_{-}K_{m-\tau}(R\alpha_{-})]\\
  0 &   -J_{m}(R\mu_{+})& \Re[c_+K_{m}(R\alpha_{+})]& \Im[c_{-}K_{m}(R\alpha_{-})] \\
  0 & -\tau J_{m+\tau}(R\mu_{+})& \Re[d_+K_{m+\tau}(R\alpha_{+})]& \Im[d_{-}K_{m+\tau}(R\alpha_{-}) \\
\end{array}%
\right)\left(
\begin{array}{cccc}
  D_{1}^{\tau}\\
   D_{2}^{\tau}\\
    C_{1}^{\tau}\\
  C_{2}^{\tau}
\end{array}
\right)=0,
\end{equation}
\end{widetext}
where $b_{\pm}=\alpha_{\pm}/(E-\tau\delta)$, $c_{\pm}=\left[ (E-\tau \delta)^2+\alpha_{\pm}^2 \right]/(E-\tau\delta)$, and $d_{\pm}=\alpha_{\pm}\left[ (E-\tau \delta)^2+\alpha_{\pm}^2 \right]/(E^{2}-\delta^{2}).$ The energy levels $E_{m,n}(R)$ of a GQB with radius $R$ can be found through the roots of the determinant of the matrix $\mathbf{M^{\tau}}$. Here, $n$ is the radial quantum number corresponding   to  $\vert n \vert$ modes in the radial direction that emerge  with increasing the size of the blister. Subsequently, one can obtain the corresponding wavefunction by solving at the given energy and size $R$ for the coefficients $C^{\tau}_{i}$ and $D^{\tau}_{i}$ and obtaining the eigenwavefunction $\Phi^{\tau}_{m,n}(\vec{r})$. From this, the radial probability density (RPD) can be found as\cite{Cohnitz2016,Lain1981}  %
\begin{equation}
\mathcal{P}_{m,n}^{\tau}(\rho) = \rho \left\vert \Phi_{m,n}^{\tau}(\vec{r})\right\vert^{2}~.
\label{probability_density}
\end{equation}
Finally, the local density of states $\mathcal{D}(\vec{r},E)$ for a GQB with radius $R$ can be derived from the eigenstates as
\begin{equation}
\mathcal{D}(\vec{r},E) = \sum_{m,n} \delta(E - E_{m,n}) \left\vert \Phi_{m,n}^{\tau}(\vec{r})\right\vert^{2}~.
\end{equation}
In the numerical results displayed in the following section we will replace the Dirac function by a Gaussian profile with a finite spectral width $\Gamma$ \cite{Van_Duppen01_2013}. 

\section{Confined states in a GQB}\label{Results}
\subsection{Homogeneous inter-layer bias}
 \begin{figure}[t!]
\vspace{0.cm}
\centering\graphicspath{{./Figures/}}
\includegraphics[width=3.4  in]{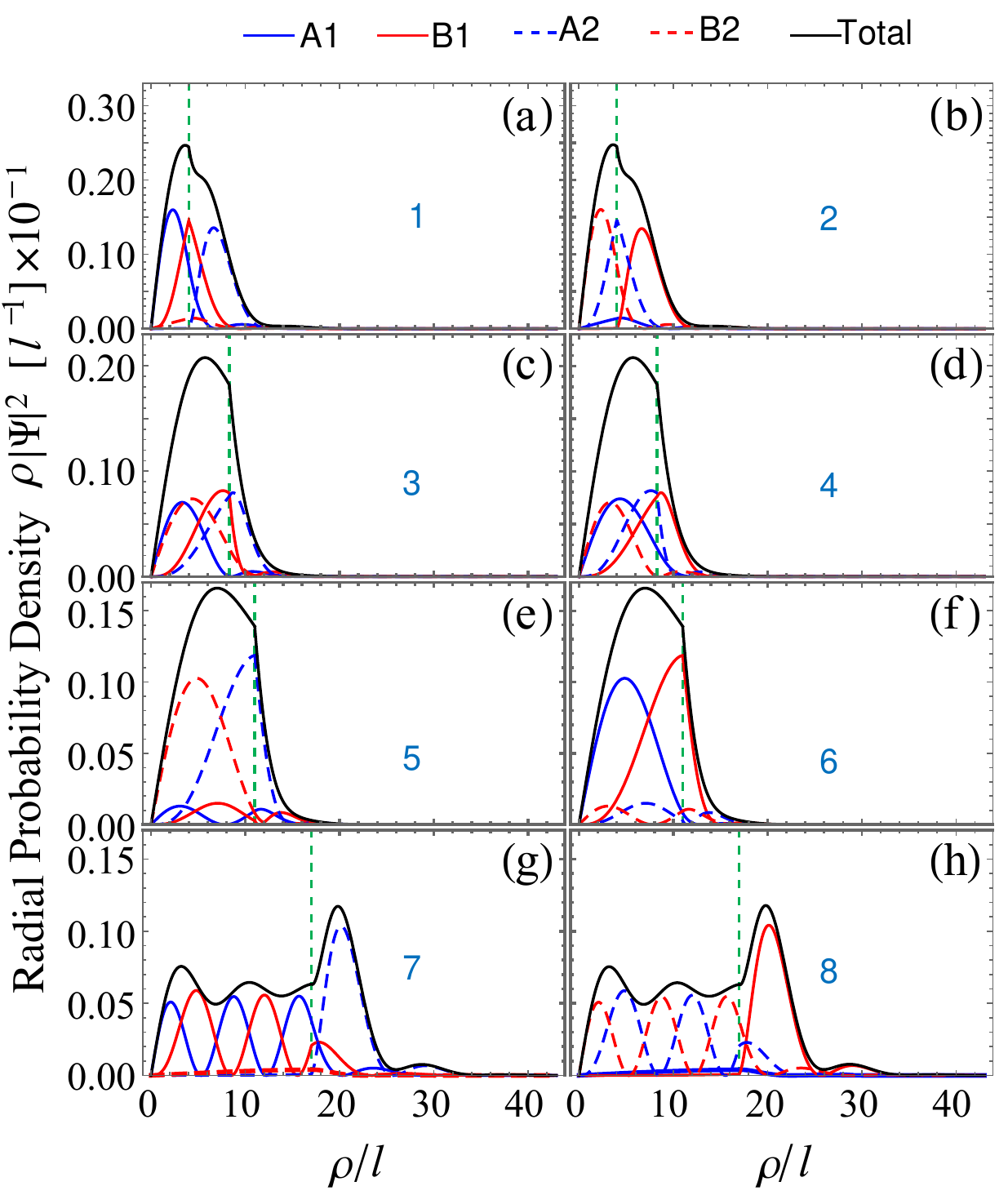}
\caption{  (a-h) Radial  probability density of $m=0$ stats in Fig. \ref{GQB_energy_levels_same_bias}(a) labelled by (1-8), respectively. The green dashed vertical line represents the radius of the GQB. Note that the states close to the continuum spectrum are mainly localized outside the GQB and preferably on the disconnected sublattices A2 and B1 as indicated in panels (g) and (h), respectively.}\label{Radial_Probability_Density_same_bias}
\end{figure}

\begin{figure}[t!]
\vspace{0.cm}
\centering\graphicspath{{./Figures/}}
\includegraphics[width=\linewidth]{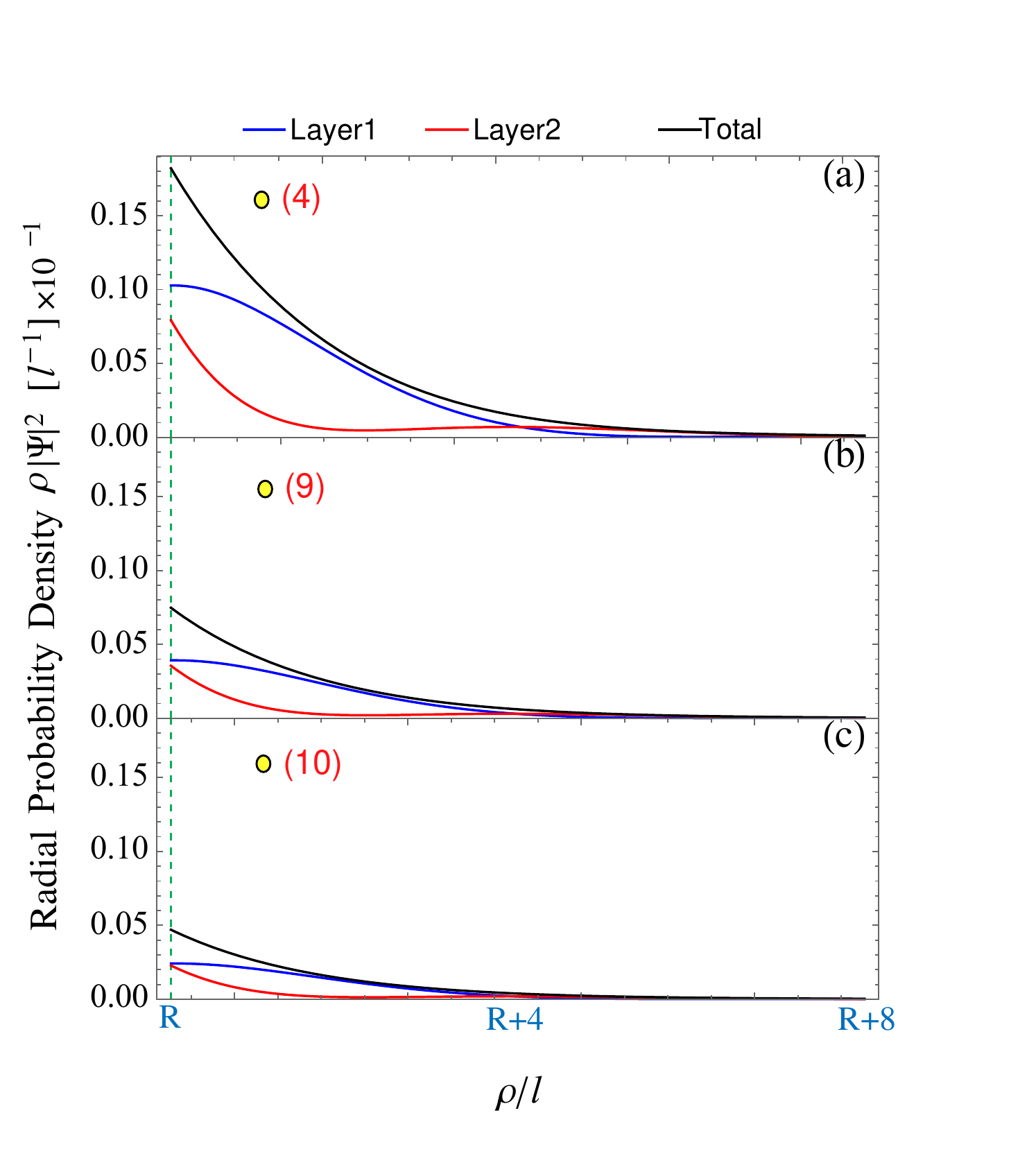}   

\vspace{0.cm}
\caption{Radial  probability density outside the blister of  the states $E_{0,1}$  in Fig. \ref{GQB_energy_levels_same_bias}(a) labelled by yellow points (4,9,10). The green vertical line represents the radius of the GQB.  } \label{RPD_Outside}
\end{figure}
Now we turn to the discussion of the numerical results for the energy levels in three configurations of GQBs. Specifically,  we consider the bias inside the blister to be the same as outside, opposite, and zero. The first configuration is the simplest example, i.e. the case where a homogeneous inter-layer bias potential $\delta$ is applied to the entire system. In Fig. \ref{GQB_energy_levels_same_bias} we show the energy levels as a function of the size $R$ of the GQB for $m=0, \ldots, \pm 3$ with $\delta = 0.25~\gamma_{1}^{0}$. The results indicate that, indeed, for the energy range as defined in Eq. \eqref{eq_energy_range} the GQB has confined modes. Panel (a) of Fig. \ref{GQB_energy_levels_same_bias} shows that in the limit $R\rightarrow 0$, the GQB has two $m=0$ confined modes at energy $\pm \delta_{\rm G}$. As the radius of the GQB increases, the modes approach each other, anti-crossing one another around $E = 0$. As the radius increases further, more $m=0$ modes are allowed inside the GQB. For a given radius, these modes are denoted by the radial quantum number $n$. The number of the confined modes  crucially depends on the strength of the applied bias outside the blister $\delta_{>}$ and its size $R$.   The energy spectrum of the different modes form anti-crossings with each other. As a consequence, the energy levels oscillate with the size of the GQB. These anti-crossings are a manifestation of coupling of states.  This coupling is established through the coupled layers outside the blister, where the confined states on upper and lower layer can feel each other.

In panels (b) - (d) of Fig. \ref{GQB_energy_levels_same_bias} we show the energy levels for non-zero angular quantum number $m$. These modes are only supported at larger radii $R$ but their characteristic behavior is similar as the $m=0$ case shown in panel (a). Notice that the results are the same for positive as for negative $m$, i.e $E_{m,n}=E_{-m,n}$. This is in contrast with previous studies where the symmetry between both signs of the angular quantum number is broken.\cite{Recher2009,Costa2014,Mirzakhani2016} 

To investigate the character of the different energy levels and the behaviour at the anti-crossings, in Fig. \ref{Radial_Probability_Density_same_bias} we show the RPD for $m=0$ at different configurations as indicated by the points in Fig. \ref{GQB_energy_levels_same_bias}(a). We choose these points to be exactly at an anti-crossing,   before, and after, and also near the continuum spectrum. Furthermore, we show the contribution of each sublattice to the probability density. Comparing for instance Figs. \ref{Radial_Probability_Density_same_bias}(a) and (b),  we see that   inside the GQB  mainly holes reside on layer 1 while  electrons  reside on layer 2. In addition, we infer from Figs. \ref{Radial_Probability_Density_same_bias}(c,d) that at the anti-crossing electrons and holes are equally distributed on both layers. A transition in the residence of states takes place when passing an anti-crossing. The states with negative energy mainly reside on layer 2 instead of layer 1 before the anti-crossing point and vice versa for the positive one as shown in Figs. \ref{Radial_Probability_Density_same_bias}(e,f).   This means that the modes $E_{0,\pm1}$ {\color{red} } anti-cross each other in Fig. \ref{GQB_energy_levels_same_bias}(a) and  correspond mainly to states on one of the two layers before or after an anti-crossing. Peculiarly, however,  in this configuration we find that states close to the continuum spectrum have   also a large part of the probability  located  outside the GQB, see Figs. \ref{Radial_Probability_Density_same_bias} (g,h).
 \begin{figure}[t!]
\vspace{0.cm}
\centering\graphicspath{{./Figures/}}
\includegraphics[width=\linewidth]{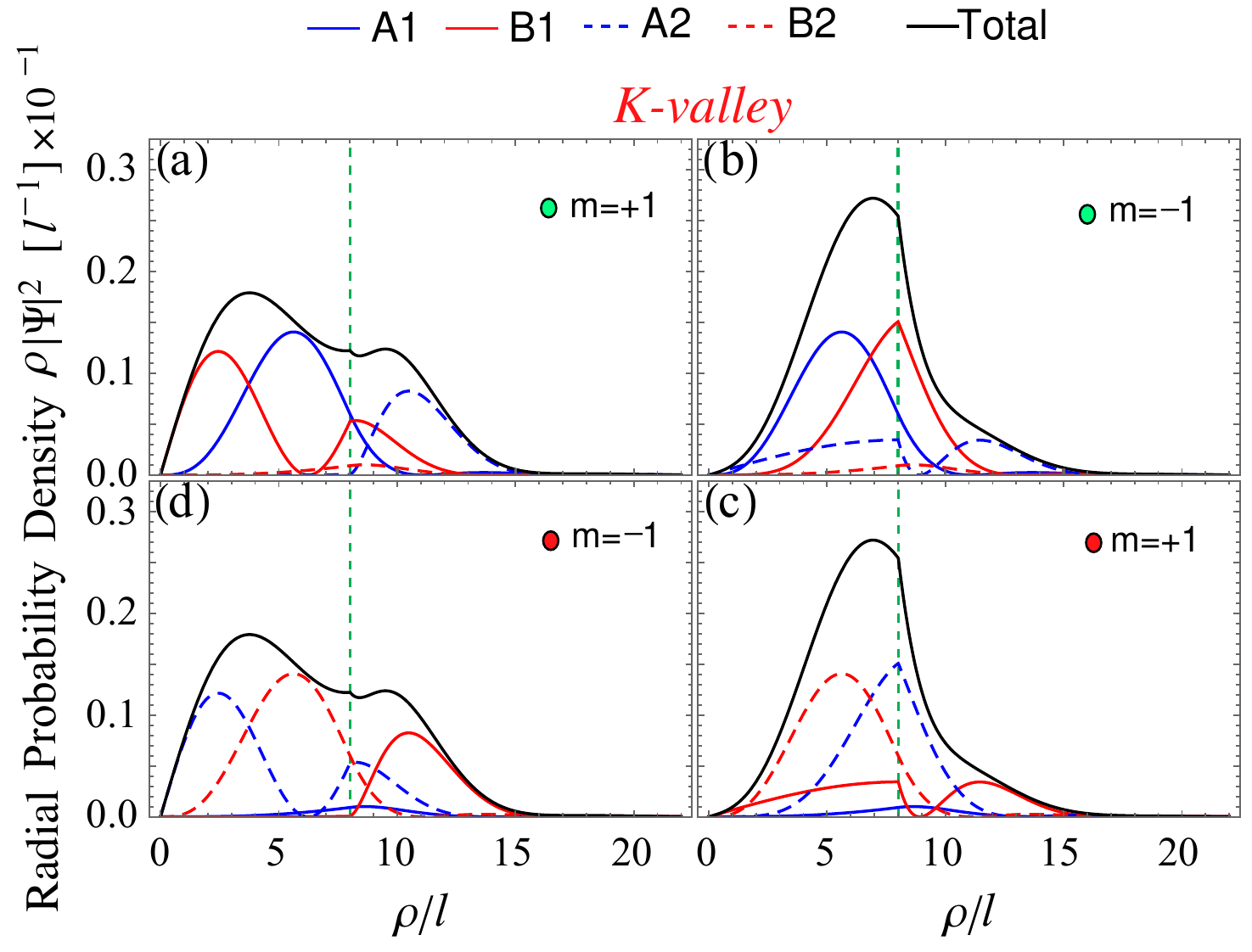}\\
\caption{The radial  probability density of $m=\pm1$ states in Fig. \ref{GQB_energy_levels_same_bias}(b) labelled by red and green dots. Top and bottom panels are  for the $K-$valley.   The green vertical line represents the radius of the GQB.  Note that the layer RPD in the vicinity of the $K'-$valley is connected through     $\left[ \mathcal{P}_{m,n}^{\tau}(\epsilon,\rho) \right]_{\rm Layer1} =\left[ \mathcal{P}_{m,n}^{-\tau}(\epsilon,\rho) \right]_{\rm Layer2}$  while  the total radial  probability density is the same in both valleys.} \label{Radial_Probability_Density_same_bias_m1}
\end{figure} 

We can find the radii $R_j$ {\color{red} } at which anti-crossing occur through the following partitioning relation:
\begin{subequations}
\begin{equation}\label{condition_anticrossing_point}
 \int_{0}^{R_{j}}d\rho\left[ \mathcal{P}_{m,n}^{\tau}(\rho) \right]_{\rm Layer 1}=\int_{0}^{R_{j}}d\rho\left[ \mathcal{P}_{m,n}^{\tau}(\rho) \right]_{\rm Layer 2},
\end{equation}
with
\begin{equation}
\left[ \mathcal{P}_{m,n}^{\tau}(\rho) \right]_{{\rm Layer}\ i}=\rho\left(\left\vert \phi^{m,n,\tau}_{Ai}(\rho) \right\vert^{2}  +\left\vert \phi^{m,n,\tau}_{Bi}(\rho) \right\vert^{2}\right)
\end{equation} 
 where the radial part of the wave functions is normalized according to
\begin{equation}\label{Normalization}
 2\pi\int_{0}^{\infty}d\rho\sum_{i=1,2}\left[ \mathcal{P}_{m,n}^{\tau}(\rho) \right]_{{\rm Layer}\ i}=1.
\end{equation}
\end{subequations}
In other words, two energy levels form an anti-crossing when  the probability of finding a state inside the GQB is the same for both layers. Note  that a point in the middle between two subsequent anti-crossings  associated with modes $E_{m,n}$  also satisfies Eq. \eqref{condition_anticrossing_point}. This point  coincides with the anti-crossing  in the second pair of energy branches, i.e. $E_{m,\left\vert n \right\vert +1}$, in the spectrum as can be inferred from Fig. \ref{GQB_energy_levels_same_bias}(a).
   
Using  Eq. \eqref{condition_anticrossing_point}, we can find the radii $R_j$  where the anti-crossings  occur for any pair of energy branches $E_{m,n}$. For example, the first three anti-crossings  of the first pair of energy branches $E_{0,\pm1}$ in Fig. \ref{GQB_energy_levels_same_bias}(a) are located at $R_{j}=(8.30,20.74, 33.27)l$. In between these three anti-crossings there are two points, where both layers also contribute the same to the RPD inside the GQB, located at $R_{j}=(14.46,26.98)l$. For large GQB we notice that the strength of the anti-crossings becomes weaker. This is a result of  leaking interaction between the two layers through the BLG outside the blister. In  Fig. \ref{RPD_Outside} we show the RPD outside the GQB at the first three anti-crossings labelled by the yellow dots (4, 9, 10) in Fig. \ref{GQB_energy_levels_same_bias}(a).  We see that the interaction between states on both layers becomes smaller with increasing the radius of GQB. Hence, in the limit $R\rightarrow \infty$ we expect the RPD to be zero outside the GQB and as result the anti-crossings will vanish and the states will be completely localized inside the blister. In this case, the GQB can be seen as a biased bilayer graphene nano-disk. We superimpose the first energy levels of a biased bilayer nono-disk with the respective angular momentum as  gray dashed curves on Fig. \ref{GQB_energy_levels_same_bias}. For a bilayer nano-disk we implement hard wall boundary conditions and the energy levels can be found by solving
 \begin{equation}
J_{m}(\mu_{+} R)=0, \ J_{m}(\mu_{-} R)=0
\end{equation}
The asymptotic behaviors of bessel function for small and large argument are
\begin{equation}\label{asympt_Bessel0}
J_{0}(x)=\begin{cases}\sqrt{2/\pi x} \cos\left( x-\pi/4 \right)& x\rightarrow \infty\\1-x^{2}/4  &x\rightarrow 0
\end{cases},
\end{equation}
using Eq.\eqref{asympt_Bessel0} one can show that $E\sim1/R$ for small and large size of bilayer nano-disk.

For $m\neq0$, we investigate the contribution of the two layers to the probability density for only $m=\pm1$ as shown in Fig. \ref{Radial_Probability_Density_same_bias_m1} , and the findings also apply for $\left\vert m \right\vert>1$. We choose two points before the first anti-crossing  of the modes $E_{1,\pm1}$ marked by red and green dots in Fig. \ref{GQB_energy_levels_same_bias}(b). Because of the symmetry between $m$ and $-m$ in this case $(E_{m,n}=E_{-m,n})$, these two points correspond to four modes as indicated in Fig. \ref{Radial_Probability_Density_same_bias_m1} .We see that    the states $m=\pm1$ with negative energy   (green dot) mainly reside on the lower layer and vice versa for states with positive energy. This corresponds to the case with $m=0$ and it also holds here for the modes $E_{\pm1,\left\vert n \right\vert>1}$. Similar to the spectrum of $m=0$, the radii for which the energy forms anti-crossings can be also found using Eq. \eqref{condition_anticrossing_point}. Notice that from the top panel of  Fig. \ref{Radial_Probability_Density_same_bias_m1} the RPD acquires the  layer  symmetry
\begin{subequations} 
\begin{equation}
\left[ \mathcal{P}_{m,n}^{\tau}(E,\rho) \right]_{\rm Layer1} =\left[ \mathcal{P}_{-m,n}^{\tau}(-E,\rho) \right]_{\rm Layer2}.
\end{equation}
In the bottom panel of Fig. \ref{Radial_Probability_Density_same_bias_m1}, we  show the same results as in the top panel but in the vicinity of the $K'-$valley. Comparing top and bottom panels of Fig. \ref{Radial_Probability_Density_same_bias_m1}, we find that the RPD also attains the following symmetry
\begin{equation}
\left[ \mathcal{P}_{m,n}^{\tau}(E,\rho) \right]_{\rm Layer1} =\left[ \mathcal{P}_{m,n}^{-\tau}(E,\rho) \right]_{\rm Layer2},
\end{equation}  
\begin{equation}
\left[ \mathcal{P}_{m,n}^{\tau}(E,\rho) \right]_{{\rm Layer}\ i} =\left[ \mathcal{P}_{-m,n}^{-\tau}(-E,\rho) \right]_{{\rm Layer}\ i}.
\end{equation}
\end{subequations}
Note that even though  the RPD of each layer is different in each valley, the total RPD is the same in both valleys.

\begin{figure}[t!]
\vspace{0.cm}
\centering\graphicspath{{./Figures/}}
\includegraphics[width=\linewidth]{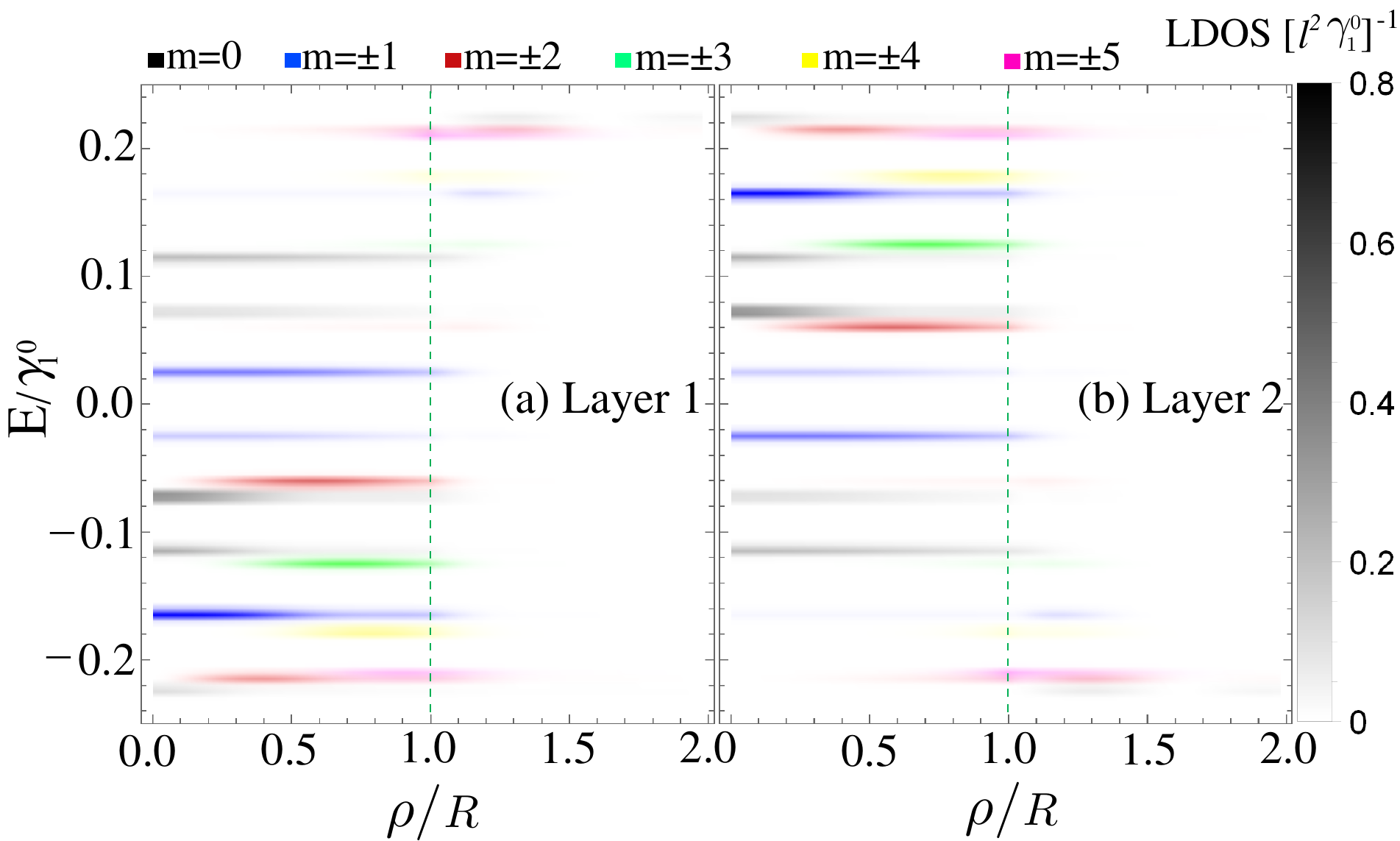}\
\vspace{0.cm}
\caption{  LDOS of the GQB for layer 1 (left) and layer 2 (right), with $R=15l$ and  $\delta_{<}=\delta_{>}=0.25\gamma_1^0$. The  spectral width of the Gaussian profile is  $\Gamma=0.02\delta_{>}$. The dashed  green  vertical lines represent the radius of the GQB.}  \label{LDOS_vs_rho_E}
\end{figure}

In Fig. \ref{LDOS_vs_rho_E} we show the local density of states for a GQB of fixed size $R=15~l$ as a function of the energy and distance from the origin for both layers. The results show that the layer selectivity of the modes is not only present for $m=0$, but also for the higher angular quantum numbers. Very pronounced is for example the $m=\pm 1$ mode that is strongly localized on layer 1 for negative energy and on layer 2 for positive energy. Such tunable layer
localization was recently also observed on topological states in
AB-BA domain walls in bilayer graphene\cite{Jaskolski2018}. Furthermore, the LDOS also shows that states with $\vert m \vert >1$ are not positioned at the center of the GQB, but more towards the edge or even outside the GQB in a classically forbidden region, specially, for those states close to the continuum spectrum. For example, the modes $\epsilon_{1,\pm5}$ exist near the continuum spectrum of the AB-BLG and meanly localized outside the blister that is about $R/3$ far from the blister's edge. Note that the closed the modes to  the continuum the far localized from the edge outside the blister.  

\subsection{Non-homogeneous inter-layer bias}
In the previous section we have considered the most straightforward case in which the inter-layer bias is the same in all parts of the sample. However, if the blister is formed by encapsulation of a metal colloid, the applied electric field also induces a dipole in the metallic nanoparticle. This will change the electrostatic potential on each layer. As a result, it can be strongly reduced inside the GQB with respect to outside it. To incorporate this difference, we investigate the case for which the inter-layer bias inside the GQB, $\delta_{<}$, is smaller than the bias $\delta_{>}$ outside. 

\begin{figure}[t!]
\vspace{0.cm}
\centering\graphicspath{{./Figures/}}
\includegraphics[width=\linewidth]{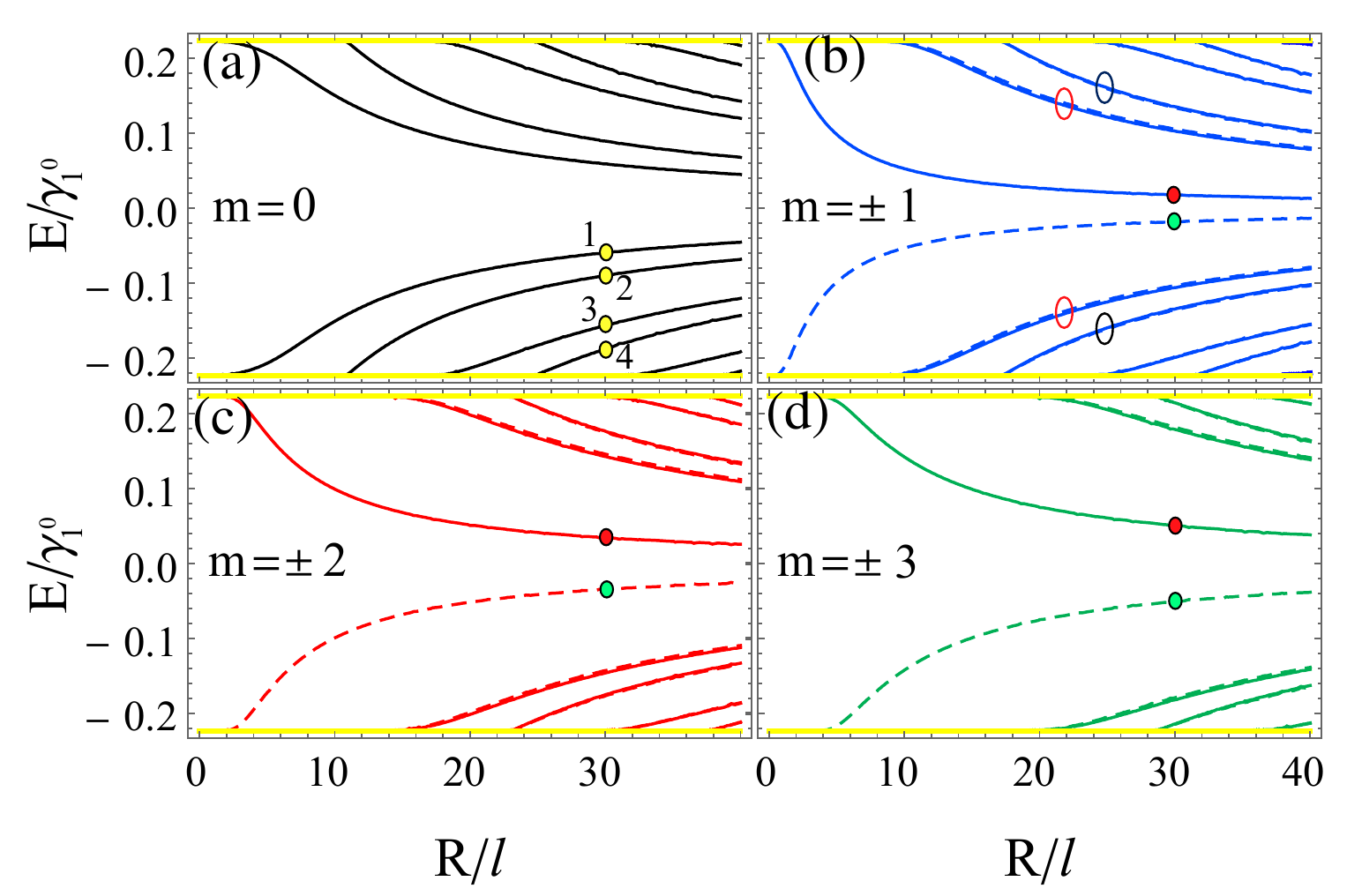}\
\vspace{0.cm}
\caption{Energy levels of the GQB as a function of its radius with  different bias inside and outside the GQB  $\delta_{<}=0$ and $ \delta_{>}=0.25 \gamma^0_{1}$. Solid (dashed) curves are for $m > 0$ ($m < 0$) where yellow horizonal lines delimit the  gap outside the GQB.  } \label{GQB_energy_levels_zero_bias}
\end{figure}

\begin{figure}[t!]
\vspace{0.cm}
\centering\graphicspath{{./Figures/}}
\includegraphics[width=\linewidth]{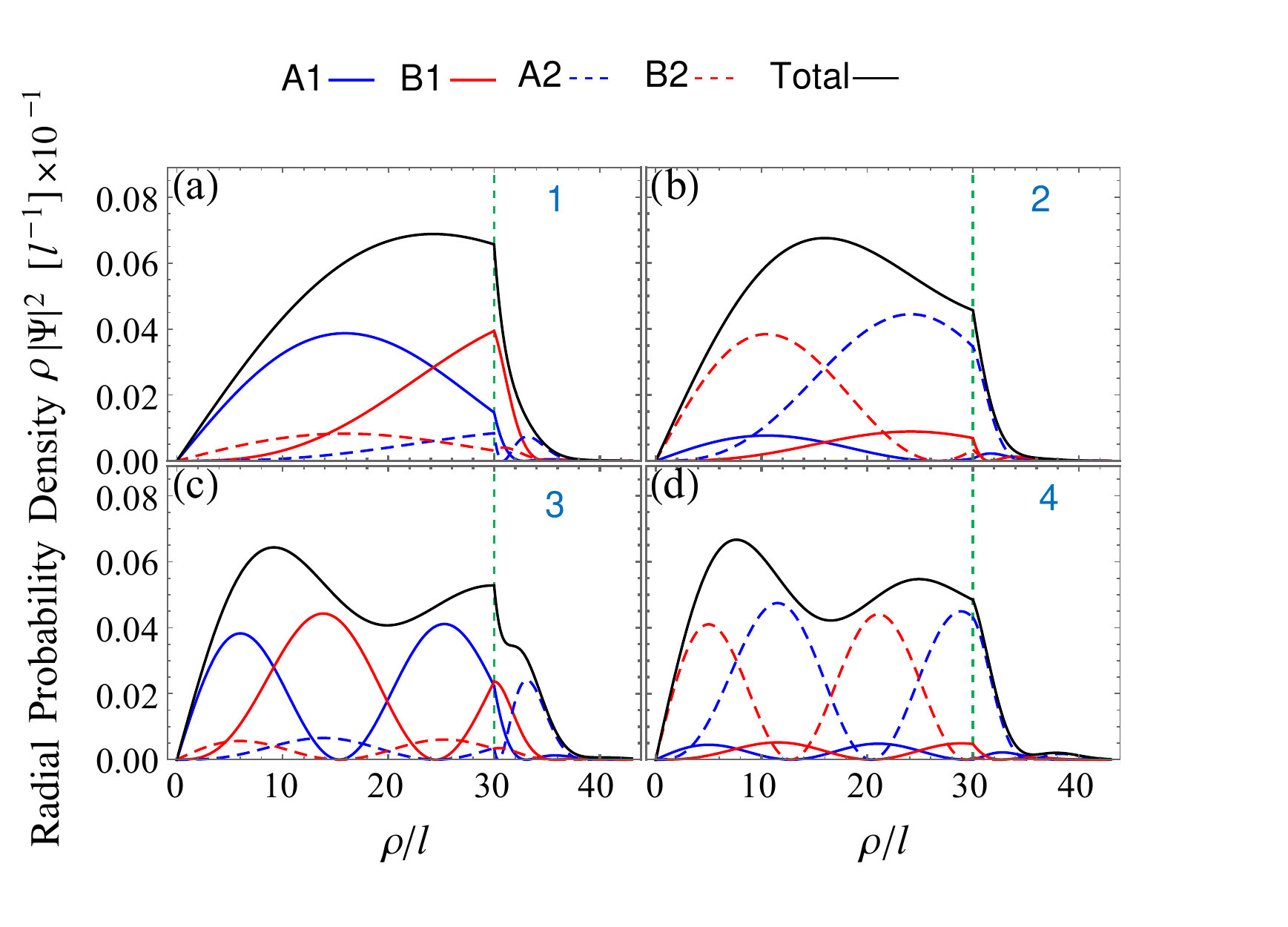}
\caption{  (a-d) Radial  probability density of the $m=0$ states in Fig. \ref{GQB_energy_levels_zero_bias}(a) labelled by (1-4), respectively. The green vertical line represents the radius $R=30l$ of the GQB. }\label{Radial_Probability_Density_zer_bias_m0}
\end{figure}

In Fig. \ref{GQB_energy_levels_zero_bias} we show the energy levels of a GQB with vanishing $\delta_{<}$ inside as a function of the radius of the blister for different values of the angular quantum number $m$. In this case, the energy levels do not show anti-crossings and approach each other as the size of the GQB increases in a monotonous way. In Fig. \ref{GQB_energy_levels_zero_bias}(a) we show the energy spectrum of the state with zero angular momentum. Here, contrary to the homogeneous bias case, each energy branch corresponds to states residing on a specific layer inside the GQB for any $R$. In Fig. \ref{Radial_Probability_Density_zer_bias_m0}  we show the RPD for different  energy branches, labelled by yellow dots in Fig. \ref{GQB_energy_levels_zero_bias}(a). States on the first energy branch $E_{0,-1}$ reside on the lower layer, see Fig. \ref{Radial_Probability_Density_zer_bias_m0}(a). While for the second branch $E_{0,-2}$, the states  along it reside on the upper layer as shown in Fig. \ref{Radial_Probability_Density_zer_bias_m0}(b). Similarly the third $E_{0,-3}$ and fourth $E_{0,-4}$ branches, marked by points 3 and 4 in Fig. \ref{GQB_energy_levels_zero_bias}(a),  the states reside mainly on the lower and upper layer respectively. This is illustrated in Figs. \ref{Radial_Probability_Density_zer_bias_m0}(c,d). Note that for the counterpart  branches in the positive energy regime, the modes  residence is  opposite   compared to the negative energy branches. 

\begin{figure}[t!]
\vspace{0.cm}
\centering\graphicspath{{./Figures/}}
\includegraphics[width=\linewidth  ]{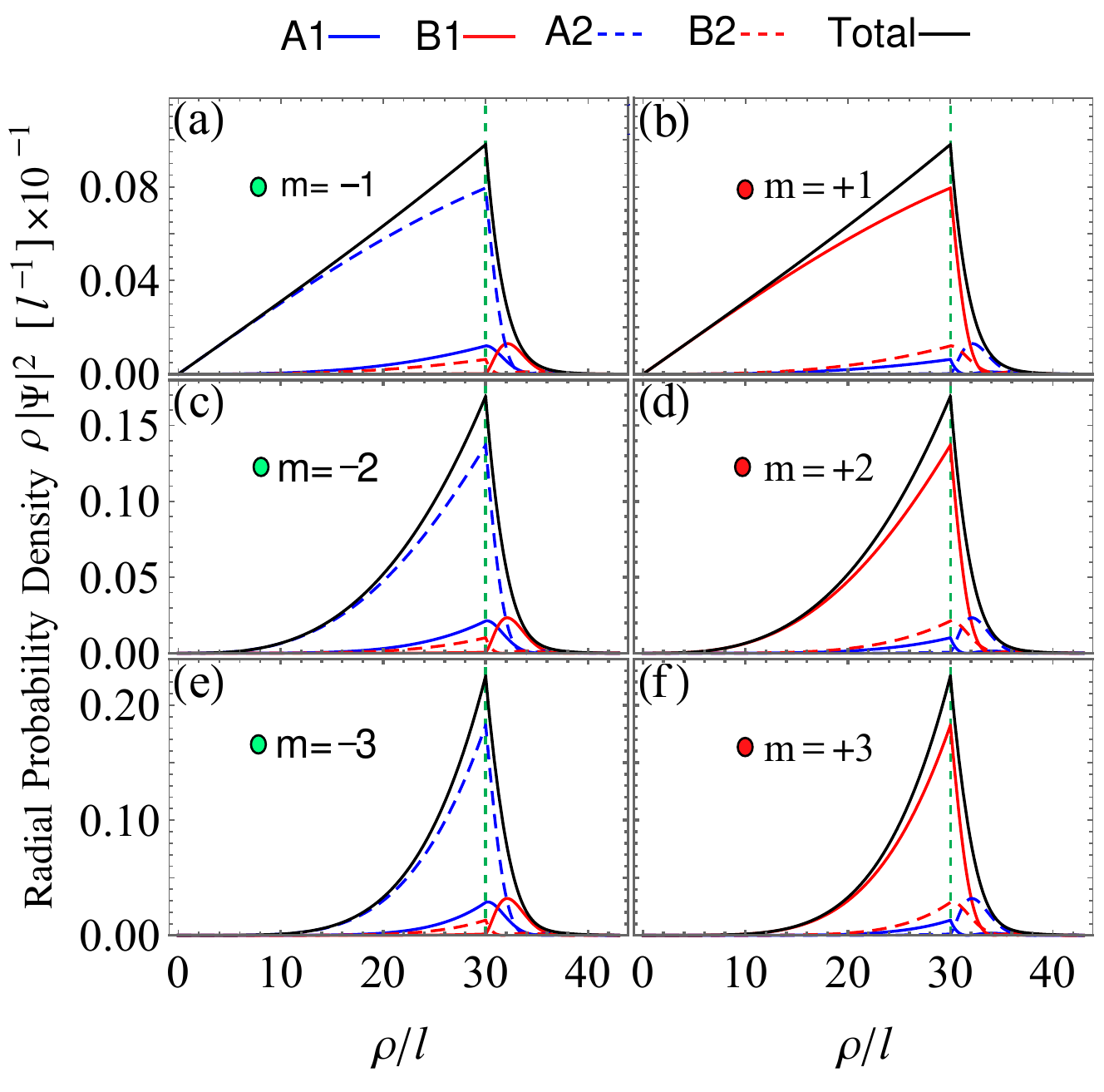}
\caption{  (a-f) Radial  probability density of $m=\pm(1,2,3)$ states in Fig. \ref{GQB_energy_levels_zero_bias}(b,c,d) labelled by green and red dots, respectively. The green vertical line represents the radius of the GQB $R=30l$. }{\color{green} }          \label{Radial_Probability_Density_zero_bias_m123}
\end{figure}

For non-zero values of the angular quantum number, however, the symmetry between positive and negative $m$ is broken. This is most clear in the first pair of modes $E_{m>0,\pm1}$ in panels (b) - (d) in Fig. \ref{GQB_energy_levels_zero_bias}. The lowest of the two is only possible for negative $m$ modes, while the highest is for positive $m$. These two modes, labelled by green and red dots in Figs. \ref{GQB_energy_levels_zero_bias}(b-d),  are mainly localized at the interface of the blister as can be seen in Fig. \ref{Radial_Probability_Density_zero_bias_m123}. Moreover,  inside the  blister they mainly reside on the upper and lower layer for $m<0$ and $m>0$, respectively, as shown in Fig. \ref{Radial_Probability_Density_zero_bias_m123}.  It turns out that they significantly reside on the disconnected sublattices A2 and B1 for the negative and positive angular momentum, respectively,  as shown in Figs. \ref{Radial_Probability_Density_zero_bias_m123}(a-f). The appearance of these localized modes at the interface of the blister is one of its quintessential traits. 

Reminding ourselves that the lowest modes mainly reside on the upper layer while the upper modes reside on the lower layer, it follows that small GQBs only can support modes with a positive angular momentum on the lower layer while the negative angular momentum-modes reside on the upper layer. For large $R$, we notice that the broken symmetry between the negative and positive angular momentum is almost restored for modes whose   radial quantum number $\left\vert n \right\vert >1$ such that $E_{m,n}(R) \approx E_{-m,n}(R)$ as shown in Figs. \ref{GQB_energy_levels_zero_bias}(b-d). The contribution of each layer to these modes is exactly the same as in the case of $m=0$ discussed in Fig. \ref{GQB_energy_levels_zero_bias}(a). For example, in the case of $m=\pm 1$, the first  pair of modes  labelled by the red  circles in Fig. \ref{GQB_energy_levels_zero_bias}(b), mainly reside on the lower  and upper layer for modes whose energy is negative and positive, respectively. The opposite  occurs for the second pair of modes, labelled by the black  circles, and  such trend also holds for  $\left\vert m \right\vert>1.$   

\begin{figure}[t!]
\vspace{0.cm}
\centering\graphicspath{{./Figures/}}
\includegraphics[width=\linewidth]{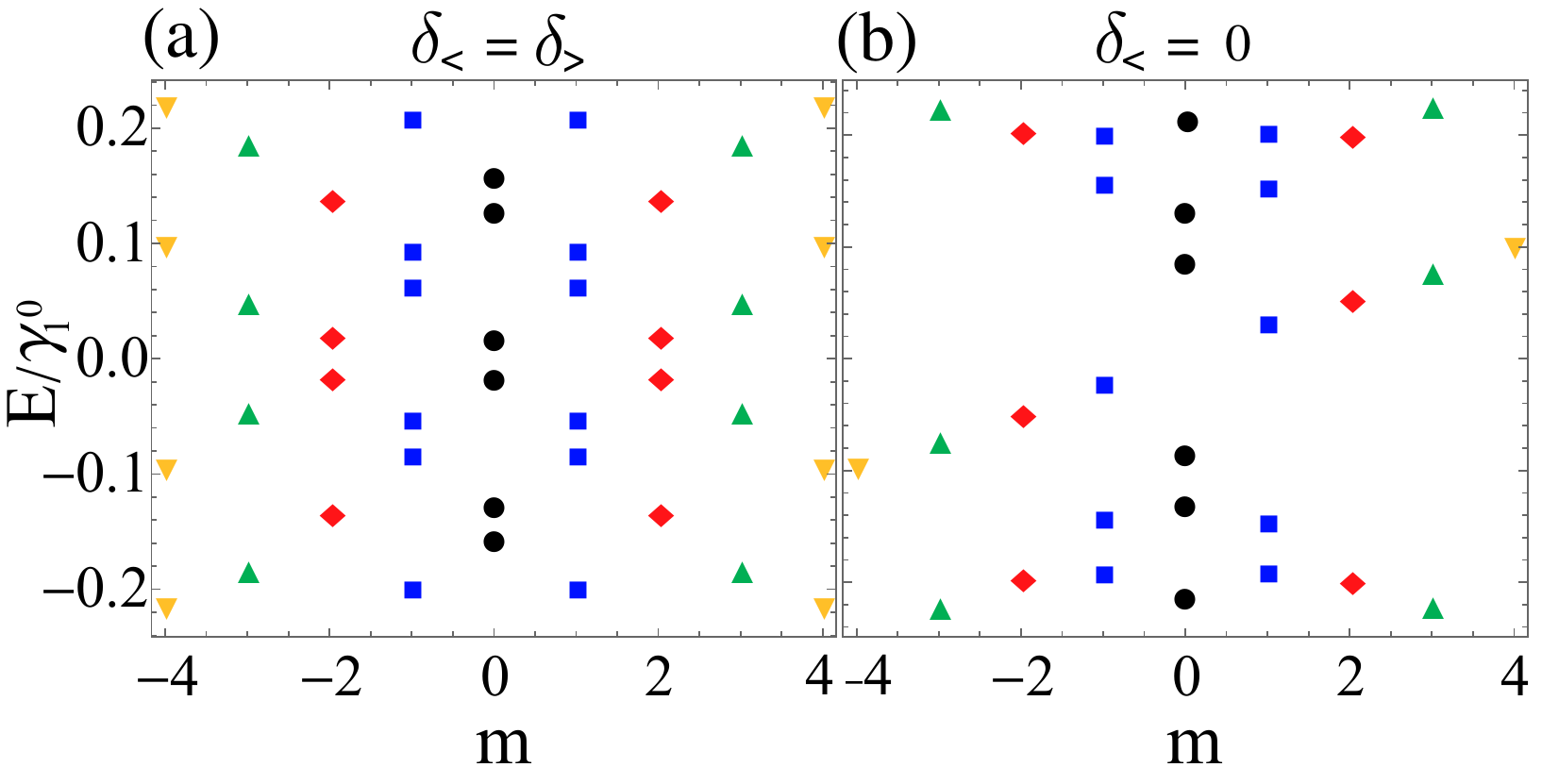}\
\vspace{0.cm}
\caption{ Energy levels of a GQB  as a
function of angular momentum label m for $R=20l$   and $\delta_{>}=0.25\gamma_{1}^{0}$ and for different values of $\delta_{<}$. } \label{energy_levels_angular_depend}
\end{figure}

In general, the energy levels still retain the following symmetry\cite{Xavier2010} 
\begin{equation}
E_{m,n}(R) = -E_{-m,n}(R)~.
\label{overall_symmetry}
\end{equation}
That this relation holds can be seen  in Fig. \ref{energy_levels_angular_depend} where the energy levels in a GQB with radius $R=20~l$ are plotted for a homogeneous inter-layer bias (panel (a)) compared with the case for a vanishing inter-layer bias (panel (b)) in the GQB. The results show that by changing the inter-layer bias inside the GQB, the $m=0$ modes are pushed away from each other while for the modes with finite angular momentum it even allows the total number of states in the GQB to be reduced. 

Note that considering inter-layer bias inside the blister that  is smaller than outside and finite also  allows confinement.   The confined modes  in this case still posses  anti-crossings  but  at larger radius $R$.
It also shows that for non-zero angular momentum the symmetry with respect to the sign of $m$ remains strongly broken for small $R$, but that the modes with opposite $m$ make a transition between different radial quantum number $n$.

\begin{figure}[t!]
\vspace{0.cm}
\centering\graphicspath{{./Figures/}}
\includegraphics[width=\linewidth]{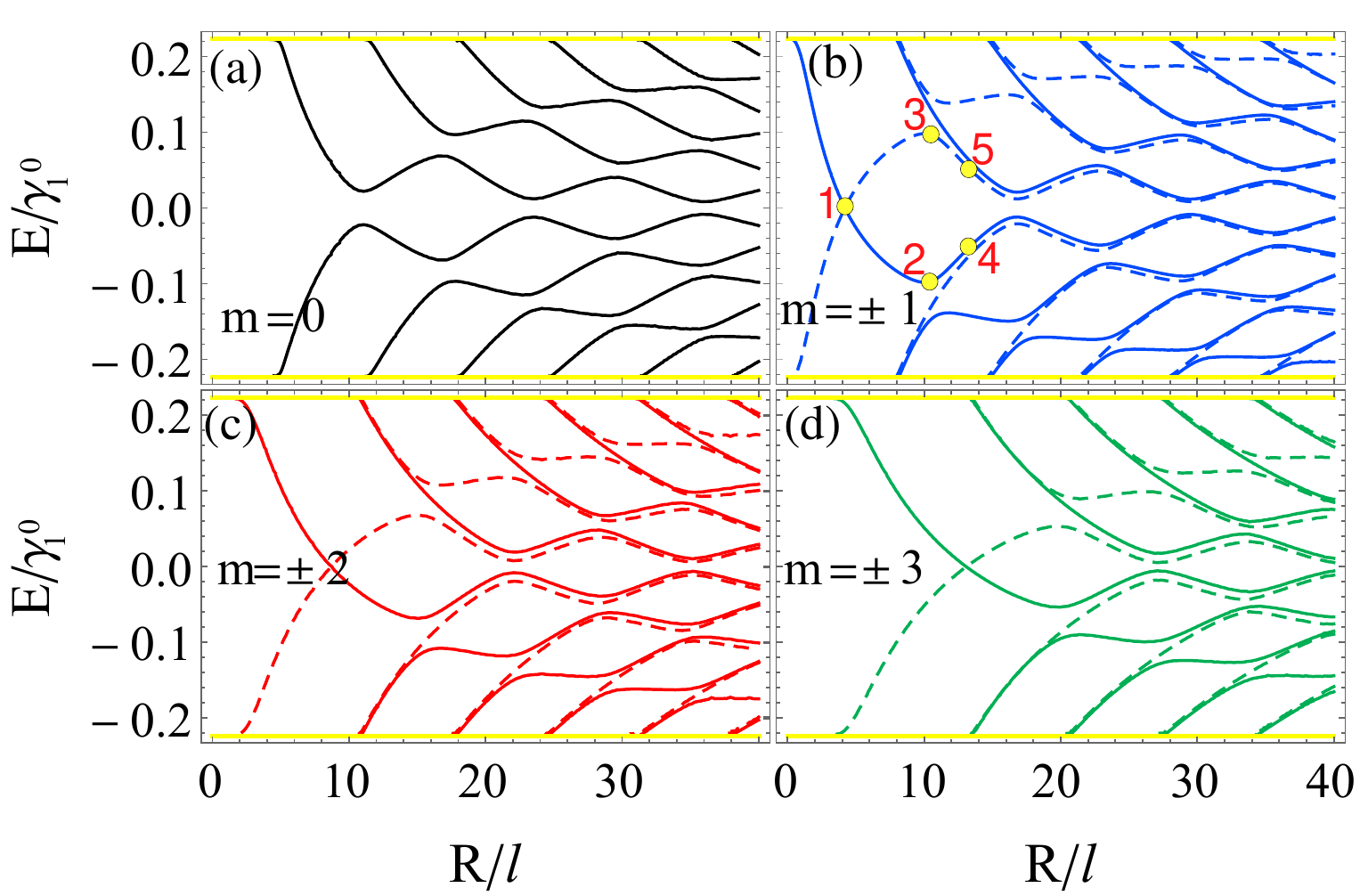}\
\vspace{0.cm}
\caption{Energy levels of the GQB as a function of its radius opposite  bias inside and outside  the GQB with    $\delta_{>}=-\delta_{<}=0.25 \gamma^{0}_1$. Solid (dashed) curves are for $m > 0$ ($m < 0$) where yellow horizonal lines delimit the  gap outside the GQB. \textcolor{blue}{}} \label{GQB_energy_levels_diff_bias}
\end{figure}

Finally, in Fig. \ref{GQB_energy_levels_diff_bias} we show the energy levels in the case when the bias inside the GQB is opposite to outside, i.e. $\delta_{<} = -\delta_{>}$. The $m=0$ result looks similar to the case of a homogeneous bias, but they are slightly  different, i.e. $E_{0,n}(R,\delta_{<}) \neq E_{0,n}(R,-\delta_{<})$. For example, the first three anti-crossing points associated with  the first pair of energy branches  $E_{0,\pm1}$   are located at $R_{j}=(10.93,23.42,35.96)l$. The anti-crossings occur for slightly larger GQB in comparison with the homogeneous inter-layer bias case. Of particular importance is also the layer residency of states, where before the anti-crossing the states with negative energy mainly  resided  on the upper layer inside the GQB and vice versa for states with  positive energy. This can be understood by considering the blister to be pure electron or hole doped\cite{Abdullah2018}.  Note that this is exactly the  opposite to what happened in the case with homogeneous bias.

\begin{figure}[t!]
\vspace{0.cm}
\centering\graphicspath{{./Figures/}}
\includegraphics[width=\linewidth]{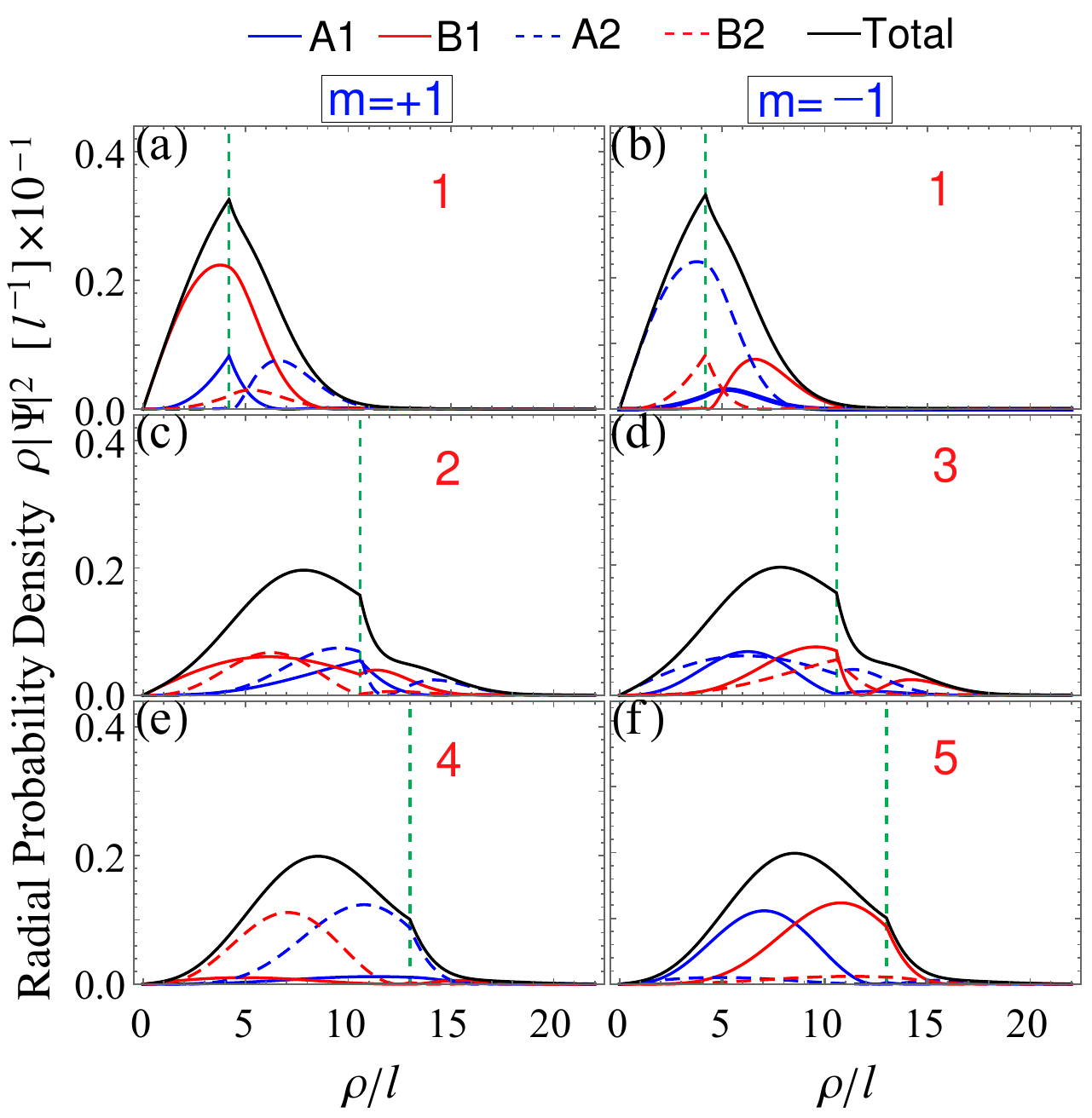}
\caption{ (a-f) Radial  probability density of  states $\epsilon_{\pm1,\pm1}$  in Fig. \ref{GQB_energy_levels_diff_bias}(b) labelled by points (1-5). The green vertical line represents the radius of the GQB. {\color{green} }          } \label{Radial_Probability_Density_opp_bias_m1}
\end{figure}

For a finite angular momentum, the results in Fig. \ref{GQB_energy_levels_diff_bias} show that the modes get pushed into each other, forming anti-crossings when the angular momentum quantum number is the same, while it crosses when $m$ is opposite. As a result, for a given inter-layer bias, there is a radius $R$ for which each of the non-zero $m$ modes are degenerate at $E_{m,1} = -E_{-m,1} = 0$. We notice that the results of non-homogeneous bias also   attains the symmetry $E_{m,n}(R) = -E_{-m,n}(R)$. In Fig. \ref{Radial_Probability_Density_opp_bias_m1}, we show the RPD of the  modes $\epsilon_{1,1}$ and $E_{-1,1}$  at different radii indicated by the yellow dots in Fig. \ref{GQB_energy_levels_diff_bias}(b). It is evident from Figs. \ref{Radial_Probability_Density_opp_bias_m1}(a,b), that the degenerated modes $E_{\pm1,1}=0$, labelled by point 1 in Fig. \ref{GQB_energy_levels_diff_bias}(b), are mainly localized at the interface of the blister.  These modes are primarily confined on the upper and lower layer inside the GQB for positive and negative angular momentum, respectively, as can be seen from Figs. \ref{Radial_Probability_Density_opp_bias_m1}(a,b). While at points 2 and 3 the two layers contribute exactly the same to the confinement of the two states $ E_{\pm1,1}=\mp0.1 \gamma_{1}^{0} $ as shown in Figs. \ref{Radial_Probability_Density_opp_bias_m1}(c,d). Then, a transition occurs in the layer confinement, where at the points 4 and 5 the states  $ E_{\pm1,1}=\mp0.06 \gamma_{1}^{0} $  principally reside on the upper and lower layer for positive and negative angular momentum, respectively. The behaviour of  the  rest of modes $E_{\pm1,\left\vert n \right\vert>1}$ resembles  that of   the case $m=0$ in Fig.  \ref{GQB_energy_levels_diff_bias}(a). This means that before and after anti-crossings, the states with negative and positive angular momentum  reside on the upper and lower layer, respectively, while at the anti-crossings they are equally distributed. Analogously for $\left\vert m \right\vert>1$, we find that modes  behave similarly to the case of $m=\pm1$.

\subsection{Effect of the inter-layer coupling}
The inter-layer coupling $\gamma_{1}$ inside the GQB decreases very fast as the height of the blister  increases However, if the blister is small and the layers in the blister remain loosely coupled, it is expected to show a band gap that is much smaller than outside the blister. Therefore, one also expects to find confined modes in this case. In this section, we investigate the energy levels of a non-zero inter-layer coupling in a GQB.
 In general, for a fixed gap outside the blister, the number of anti-crossings and their location  mainly depend on  the bias inside the blister. This  allows to control the confinement to be mainly localized on a specific layer. 

\begin{figure}[t!]
\vspace{0.cm}
\centering\graphicspath{{./Figures/}}
\includegraphics[width=\linewidth]{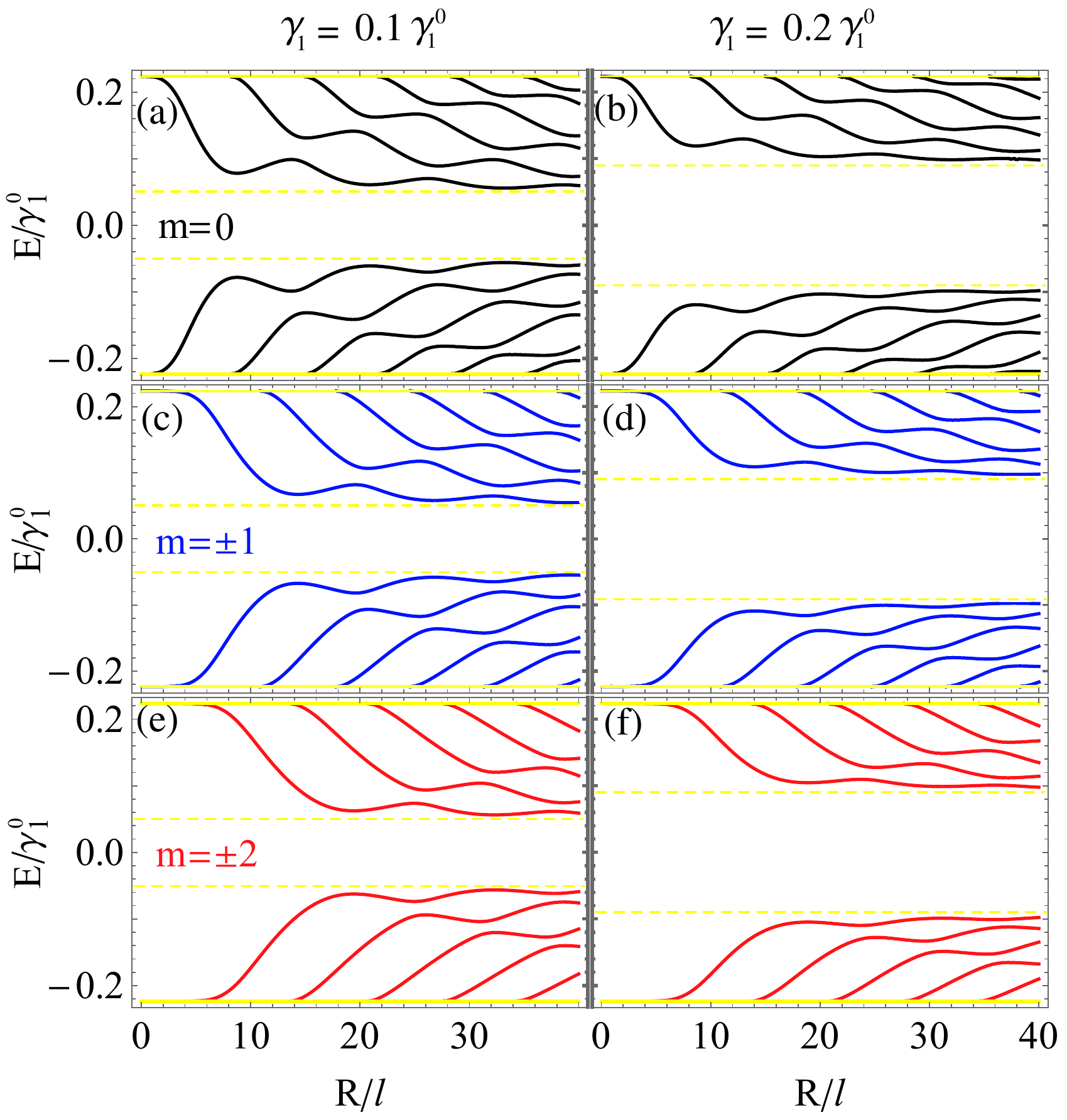} \caption{Energy levels of GQB  with different strength of the inter-layer coupling inside it, where yellow-horizonal dashed and solid   lines delimit the  gap  inside and outside the GQB, respectively, with  $\delta_{<}=\delta_{>}=0.25 \gamma_{1}^{0}$.} \label{GQB_energy_levels_bias-diff_gamma1}
\end{figure}

In Fig. \ref{GQB_energy_levels_bias-diff_gamma1} we show the energy levels of a GQB with a homogeneous inter-layer bias but different inter-layer coupling inside the blister as a function of the radius $R$. We see that for small radii, the energy levels are similar to the case of a completely decoupled blister, but that as the GQB grows, the energy levels do not cross the gap formed inside the GQB. As a consequence, for large $R$ the oscillations of the lowest positive energy level are decreased and this level approaches the value of $\delta_{\rm G}$, dashed yellow lines, from Eq. \eqref{eq_energy_range} calculated with the inter-layer coupling inside the blister.

Finally, the results in the current paper can be elegantly  verified by measuring the conductance of electrons through the GQB. According to a recent experiment\cite{Gutierrez2016}, a quantum dot with the size of few nanometers was  realized and its  electronic spectrum was probed  by  STM measurements. Thuse, The same approach can be used to prob the electronic spectrum of the GQB. The ideal setup for this purpose  is to apply a homogeneous bias to the GQB of strength that allows few modes to be confined within few nanometers size of the blister.  The regime where the electronic properties of the GQB can be amenable to STM measurements restricted to a global bias of strength in the range $\backsim(30-100)$ meV with the size $R$ of the order $\backsim (15-40)$ nm. Note that even the bias is not completely homogeneous or the inter layer coupling inside the blister is not strictly zero, confined states still exist.

\section{Conclusions}\label{Conclusions}
In this paper, we used the continuum model to analytically calculate the wave functions and  thus the discrete    energy levels of bound states  trapped in a locally delaminated bilayer graphene system  that is called GQB.  We have investigated the energy spectrum and eigenstates of such  system under the application of an electrostatic  potential difference between the graphene layers. We considered three situations where the bias inside the blister is the same as outside, opposite, and zero. The energy spectrum of GQB is layer localized where the confined states localized on either  layer or fairly distributed.    For a biased blister, as the radius of the GQB increases, the energy levels show anti-crossings when the occupation inside the GQB is the same on both layers. For small size of GQB, the localized states on each layer can feel each other through the AB-BLG flakes outside the blister leading to  strong anti-crossings that steadily    decreasing with  increasing the blister's size.   These anti-crossings disappear   in case we consider pristine  blister; however, the layer localization is still maintained. When a non-homogenous bias is applied to GQB, we found that the system can support edge modes with finite angular momentum. These modes reside mainly on one layer, while the mode with opposite angular momentum resides on the other layer. In addition, the confined modes exhibit certain symmetry that can be altered by the electrostatic bias such that $E^{(e)}_{n,m}=E^{(h)}_{n,\mp m}$ with $-(+)$ corresponds  to zero (homogenous) bias inside the blister.   

Finally, we have shown that the electronic confinement also occurs if the decoupling of both layers is not perfect. By assessing the effect of a residual decrease of inter-layer coupling inside the blister, we showed that outside the energy gap, confined states are also expected.

With this paper, we show that GQBs can form an interesting platform for new
types of graphene-based\newpage quantum  dot systems. As the creation of the dot can be solely tuned by application of a gate to a blistered system, we expect that the work in this manuscript can be elegantly  verified by STM measurements.

%%%%%%%%%%%%%%%%%%%%%%%%%%%%%%%%%%%%%%%%%%%%%%%%%%%%%%%%%%%%%%%%
\section*{Acknowledgments}
%%%%%%%%%%%%%%%%%%%%%%%%%%%%%%%%%%%%%%%%%%%%%%%%%%%%%%%%%%%%%%%%
HMA and HB acknowledge  the Saudi
Center for Theoretical Physics (SCTP) for  their generous support and the support of KFUPM under physics research group projects  RG1502-1 and RG1502-2. This work is supported by the Flemish Science Foundation (FWO-Vl) by a post-doctoral fellowship   (BVD).

%%%%%%%%%%%%%%%%%%%%%%%%%%%%%%%%%%%%%%%%%%%%%%%%%%%%%%%%%%%%%%%%
%%%%%%%%%%%%%%%%%%%%%%%%%%%%%%%%%%%%%%%%%%%%%%%%%%%%%%%%%%%%%%%%%
%%%%%%%%%%%%%%%%%%%%%%%%%%%%%%%%%%%%%%%%%%%%%%%%%%%%%%%%%%%%%%%%%

%%%%%%%%%%%%%%%%%%%%%%%%%%%%%%%%%%%%%%%%%%%%%%%%%%%%%%%%%%%%%%%%%
%%%%%%%%%%%%%%%%%%%%%%%%%%%%%%%%%%%%%%%%%%%%%%%%%%%%%%%%%%%%%%%%%
% \bibliographystyle{physrev} 
% \bibliography{Library}

%%%%%%%%%%%%%%%%%%%%%%%%%%%%%%%%%%%%%%%%%%%%%%%%%%%%%%%%%%%%%%%%%
%%%%%%%%%%%%%%%%%%%%%%%%%%%%%%%%%%%%%%%%%%%%%%%%%%%%%%%%%%%%%%%%%

\end{document}